\documentclass[prb,twocolumn,amsmath,amssymb,notitlepage]{revtex4-1}
\usepackage{graphicx,color,hyperref}
\usepackage{bbm}

\newcommand{\ix}[1]{\ensuremath{\text{#1}}} 
\newcommand{\K}{\ix{K}} 
\newcommand{\Ret}{\ix{R}} 
\newcommand{\Adv}{\ix{A}} 
\newcommand{\res}{\ix{res}} 

\begin{document}

\title{The quench dynamics of a dissipative quantum system: a renormalization group study}

\def\affiliationname{Institut f{\"u}r Theorie der Statistischen Physik, RWTH Aachen University and 
JARA---Fundamentals of Future Information Technology, 52056 Aachen, Germany}

\author{O.~Kashuba}
\affiliation{\affiliationname}
\author{D.~M.~Kennes}
\affiliation{\affiliationname}
\author{M.~Pletyukhov}
\affiliation{\affiliationname}
\author{V.~Meden}
\affiliation{\affiliationname}
\author{H.~Schoeller}
\affiliation{\affiliationname}

\begin{abstract}
We study dissipation in a small quantum system coupled to an environment held in thermodynamic equilibrium.
The relaxation dynamics of a system subject to an abrupt quench in the parameters of the underlying 
Hamiltonian is investigated using two complementary renormalization group approaches.  The 
methods are applied to the Ohmic spin-boson model close to the coherent-to-incoherent transition. 
In particular, the role of non-Markovian memory for the relaxation before and after the quench 
of the spin-boson coupling and the Zeeman splitting of the up and down spin is investigated.
\end{abstract}


\date{\today}

\maketitle

\section{Introduction}
\label{sec:intro}

Obtaining a detailed understanding of a dissipation in small quantum systems coupled to a reservoir 
held in thermodynamic equilibrium poses a formidable challenge.  With the increasing accuracy of 
experimental techniques in diverse fields such as condensed matter physics, quantum optics, cold atomic 
gases, physical chemistry, and quantum information science, phenomenological approaches need
to be complemented by a more microscopic view. One thus has to develop appropriate microscopic 
models for dissipation as well as the corresponding quantum many-body methods to solve those with 
sufficient precision. We here report on progress with respect to the second aspect. Focusing on the 
commonly investigated spin-boson model (SBM),\cite{Leggett87,Weiss12} we propose two 
complementary renormalization group (RG) based methods which allow controlled access to 
the relaxation dynamics of the Ohmic SBM close to its coherent-to-incoherent 
transition\cite{Leggett87,Weiss12}: the real-time RG (RTRG) (Ref.~\onlinecite{Schoeller09}) and the 
functional RG (FRG) (Ref.~\onlinecite{Metzner12}). We study the relaxation dynamics under the SBM Hamiltonian out 
of an initially (at time $t=0$) prepared  product state of spin-up and the boson vacuum together
with the time evolution after a quantum quench where one of the parameters of the SBM is 
changed abruptly; we here exclusively consider temperature $T=0$. 

The time dependence of the position of a classical damped harmonic oscillator constitutes a 
possible point of reference for the functional dependence of the spin expectation value $P$ 
on time $t$ in the SBM.  An important intrinsic property of the SBM that is absent in the 
oscillator is the system's memory. Non-Markovian memory effects are a fundamental issue in 
dissipative quantum systems, in particular due to their relevance in quantum information theory.
Several ways have been proposed to characterize and measure to which degree the dynamics of
an open quantum system shows non-Markovian features, see e.g. Ref.~\onlinecite{Breuer12} for 
a recent review. Here, we define non-Markovian memory by those effects which result from 
the energy dependence of the effective kernel in Laplace space determining the time 
evolution of the local quantum system. 
Using RTRG, a generic analysis\cite{Pletyukhov10,Kashuba13} has shown that
non-Markovian dynamics results in additional exponentially decaying terms for the time evolution,
which have unexpected oscillation frequencies and decay rates compared to the leading Markovian
terms, together with pre-exponential functions containing power laws and logarithmic functions 
in the long-time limit. These effects result from the nonanalytic energy dependence of the effective
kernel, where the position of a branching point determines the exponential part,
whereas the pre-exponential function follows from the scaling behavior 
around the branching point. In special cases, like e.g. quantum critical points or reservoirs with specific
nonanalytic density of states, it may even happen that non-Markovian effects lead to a
pure power law in the long-time limit without any exponentially decaying part, as it is e.g. the
case for multi-channel Kondo models.\cite{Pletyukhov12}

Non-Markovian effects have already been reported for the Ohmic SBM. Its Hamiltonian reads as
\begin{align}
\nonumber
H_{\text{tot}} &= 
\frac{\epsilon}{2}\sigma_{z} - \frac{\Delta}{2}\sigma_{x} + \sum_{k}\omega_{k} b_{k}^\dagger b_{k} \\
&\hspace{0.5cm}
-\sum_{k} \frac{\lambda_{k}}{2} \sigma_{z} \left(  b_{k}^\dagger + b_{k}\right),
\label{eq:Hsbm}
\end{align}
where $\sigma_\eta$, $\eta=x,z$, are the Pauli matrices, and $b_k^{(\dag)}$ are bosonic ladder 
operators. A spin-$\frac{1}{2}$ with Zeeman splitting $\epsilon$ and tunneling $\Delta \geq 0$ between the two states 
is coupled by $\lambda_k$ to a reservoir of bosonic modes with dispersion $\omega_k$. Of interest is the
computation of the spin expectation value $P(t) = \langle\sigma_z\rangle(t)$ primarily 
considering the initial condition $\langle\sigma_z\rangle(0)=1$. $P(t)$ corresponds to the population 
imbalance of left and right double-well states when exploiting the analogy of the SBM to  
a particle in a double-well potential.\cite{Leggett87,Weiss12} The spin-boson 
coupling is characterized by a spectral density $J(\omega)=\sum_k \lambda_k^2 \delta(\omega-\omega_k)$ 
containing the microscopic details of the model.\cite{Leggett87} We concentrate on the extensively 
studied ohmic case when the spectral density is linear up to a cutoff $D$, e.g. 
$J(\omega)=2\alpha\omega e^{-\omega/\omega_c}$. In the special case without bias, $\epsilon=0$, 
the standard result quoted for the spin expectation value $P(t)$ is the one resulting from the 
noninteracting blip approximation (NIBA), where $P(t)$ decomposes 
into a coherent (pole) and incoherent (branch cut) contribution $P(t)=P_{\text{pole}}(t)+P_{\text{bc}}(t)$. 
In the long-time limit, $T_K t\gg 1$, the NIBA predicts \cite{Leggett87,Weiss12}
\begin{align}
\label{eq:NIBA_pole}
P_{\text{pole}}(t)\,&=\,\theta(1-2\alpha)\frac{1}{1-\alpha}e^{-{\Gamma_1^*}t}\cos(\Omega t)\,\\
\label{eq:NIBA_bc}
P_{\text{bc}}(t)\,&=\,\frac{1}{\Gamma(2\alpha-1)}\frac{1}{(T_K t)^{2-2\alpha}}\,.
\end{align}
Here, $\Gamma_1^*=T_K\sin\frac{\pi\alpha}{2(1-\alpha)}$ and $\Omega=T_K\cos\frac{\pi\alpha}{2(1-\alpha)}$ are
the characteristic scales determining the exponential time dynamics of the coherent part $P_{\text{pole}}(t)$, and
$T_K=\Delta\left(\Delta/\omega_c\right)^{\alpha/(1-\alpha)}$ is an effective tunneling rate. The
incoherent part $P_{\text{bc}}(t)$ is a term resulting purely from non-Markovian dynamcis, whereas the coherent
part $P_{\text{pole}}(t)$ is also obtained in Markov approximation but the precise values of $\Gamma_1^*$,
$\Omega$, and the prefactor are influenced by non-Markovian contributions.
For strong enough damping $\alpha>\frac{1}{2}$, the NIBA predicts correctly that the coherent
term $P_{\text{pole}}(t)$ is absent. Although the results \eqref{eq:NIBA_pole} and \eqref{eq:NIBA_bc} 
were originally predicted for all values of $\alpha$, controlled calculations for
small $\alpha$ and for $\alpha$ close to $\frac{1}{2}$ have shown that this result can not be
correct for all $\alpha$. For small $\alpha$ it has been shown recently that $P_{\text{bc}}(t)$
is also exponentially decaying with a rate 
$\Gamma_2^*\approx 2\Gamma_1^*$ (Refs.~\onlinecite{Slutskin11} and~\onlinecite{Kashuba13}) and the $O(\alpha)$ correction
to the exponent of the algebraically decaying function is absent \cite{Kashuba13}. Especially the
latter requires a controlled analysis close to branching points of the effective kernel, such
that all powers $\sim [\alpha\ln(T_K t)]^n$ are systematically taken into account. So far,
this has only been achieved by the RTRG method, where a controlled renormalization group study
is possible for the weak-coupling regime of small $\alpha$. 

For $\alpha$ close to the coherent-to-incoherent transition at $\alpha=\frac{1}{2}$, which is the 
regime of interest in this work, a systematic perturbative analysis is also possible in the
parameter $g=1-2\alpha$. For small $|g|\ll 1$ and for $\omega_c$ being the largest energy scale (scaling limit) 
the SBM can be mapped onto the interacting resonant level model (IRLM),\cite{Leggett87,Weiss12} 
which describes  a quantum dot connected to a lead featuring fermionic degrees of freedom with a 
constant density of states of bandwidth $2\omega_c$. Its Hamiltonian reads
\begin{align}
\nonumber
H_{\text{tot}}&=\epsilon\, d^\dagger d \!+\! \sum_{k}\varepsilon_{k} a_{k}^\dagger a_{k} +
 \sqrt{\frac{\Gamma_{0}}{2\pi\nu}}\sum_{k} 
(d^\dagger a_{k} \!+\! d a_{k}^\dagger)\\
& \hspace{1cm}
+ \frac{U}{2\nu} (d^\dagger d- d d^\dagger)\sum_{kk'}  :a_{k}^\dagger a_{k'}: ,
\label{eq:Hirlm}
\end{align}
where $d^{(\dagger)}$ is the fermionic dot ladder operator, $a_{k}^{(\dagger)}$ is an 
annihilation (creation) operator of a fermion in the lead with energy $\epsilon_{k}$, 
and the density of states in the lead is $\sum_{k}\delta(\varepsilon_{k}-\omega)=\nu$ for 
$\omega\ll \omega_c$. Normal ordering with respect to (w.r.t.) the reservoir equilibrium
distribution is denoted by $:\dots :$. The IRLM parameters are related to the ones of the SBM as 
$U=1-\sqrt{2\alpha}$ and $\Gamma_{0}=\Delta^{2}/\omega_c$. Our observable of interest $P(t)$ 
corresponds to $2 \left< d^\dag d \right>(t) -1$, 
with the expectation value of the dot level occupancy $\left< d^\dag d\right>(t)$ 
and the initial condition $\left< d^\dag d\right>(0)=1$. Throughout this work 
we mainly use the language of the IRLM but switch to the SBM language whenever this is useful 
to describe the underlying physics. 

Concerning the relaxation dynamics of $P(t)$ for $\alpha$ close to $\frac{1}{2}$,
is has already been shown by an improved NIBA calculation \cite{Egger97}  
that the incoherent part $P_{\text{bc}}(t)$ has to be changed to
\begin{equation}
\label{eq:improved_NIBA}
P_{\text{bc}}(t) \,=\,- g [1+ 3 \Theta(-g)] \frac{e^{-\Gamma_2^* t}}{(T_K t)^{1+|g|}} \,,
\end{equation}
i.e. it is also exponentially decaying with a
rate $\Gamma_2^*\approx \Gamma_1^*/2$, the prefactor changes to $-4g=4(2\alpha-1)$ for $\alpha>\frac{1}{2}$,
and the exponent of the algebraically decaying function was predicted to change to
$1-g=2\alpha$ for $\alpha>\frac{1}{2}$. However, a systematic analysis containing all powers 
$\sim[g\ln(T_K t)]^n$ with $g=1-2\alpha$ is still missing, i.e. the corrections in $O(g)$ to
the exponent can not be trusted so far. The predictions of improved NIBA were recently 
confirmed in a letter-style paper using RTRG and FRG for the IRLM.\cite{Kennes13b}
In addition, it was shown that the incoherent part $P_{\text{bc}}(t)$ has the form Eq.~\eqref{eq:improved_NIBA} 
for {\it exponentially large times} $(T_K t)^{|g|}\gg 1$ only and an analytical formula was 
proposed valid for all times $T_K t\gtrsim 1$. Furthermore, an improved expression for the rate 
$\Gamma_2^*$ was presented, and it was shown that the prefactor of the coherent part $P_{\text{pole}}(t)$ 
has to be changed to $2\frac{1-g}{1+g}=\frac{2\alpha}{1-\alpha}$.\cite{footnote_1} One purpose of this paper is to present more details of the results from RTRG and FRG for the relaxation dynamics 
close to the coherent-to-incoherent transition at $\alpha=\frac{1}{2}$.

From this discussion, it is obvious that the relaxation dynamics of the unbiased, Ohmic SBM 
differs from the NIBA prediction for $\alpha\ll 1$ and for $|1-2\alpha|\ll 1$ and it remains 
to be seen if similar deficits prevail for other $\alpha$.
The RTRG and FRG methods are perturbative renormalization group methods for nonequilibrium 
systems, and therefore their applicability range for a controlled analysis of the time evolution 
is restricted to the weak-coupling regime. However, in contrast to bare perturbation theory or 
the self-consistent Born approximation, as e.g. applied to the Ohmic spin-boson model at small 
$\alpha$,\cite{Slutskin11,DiVincenzo05} the RTRG and FRG are unique in the sense that they can
select systematically all logarithmically diverging terms in the band width $\omega_c$ in all orders
of perturbation theory, such that renormalized parameters and exponents of power laws for
the time evolution can be determined. Other numerical studies for the SBM using a variety of 
different methods, either being formally exact or selecting certain subclasses of
processes in all orders of perturbation theory, have found exponential dependence and
substantiated the coherent-to-incoherent crossover but pre-exponential functions 
were so far not identified unambiguously.\cite{Egger94,Anders06,Wang08,Orth10,Orth13,Reichman97}

For $\alpha$ close to $\frac{1}{2}$, one of the important characteristics to compare is the transition 
between coherent and incoherent relaxation. In accordance with the standard terminology 
in the field of dissipative quantum mechanics we speak of incoherent dynamics if 
$P(t)$ is a monotonically decaying function while we refer to coherent behavior 
if $P(t)$ is nonmonotonic. To understand the effect of the memory on this transition,
we study the system's dynamics in two cases: (1) when it relaxes with time-independent Hamiltonian out of a 
nonequilibrium product state, and its dynamics is affected by the memory collected over time, 
and (2) when in addition at time $t_{q}$ some of the Hamiltonian's parameters are changed 
abruptly so that the behavior of the system at $t>t_{q}$ is influenced by the interplay of 
the dynamics in the new regime and the memory collected before $t_{q}$. 
The first case we name the \emph{relaxation protocol}, implying that the small system and the bath held 
in thermodynamic equilibrium were decoupled initially 
[i.e., $\Gamma_{0}=U=0$ or $\Delta=0$, $\alpha=1/2$ for the SBM Hamiltonian 
Eq.~\eqref{eq:Hsbm}], and at the time $t=0$ the Hamiltonian in Eq.~\eqref{eq:Hirlm} 
with non-zero $\Gamma_{0}$ and $U$ comes into effect. The \emph{quench protocols} 
in addition to the steps of the 
relaxation protocol imply the sudden change of the Hamiltonian parameters 
$\epsilon$, $\Gamma_{0}$, and/or $U$ at time $t_{q}>0$ [in the language of the SBM 
the parameters $\epsilon$, $\Delta$, and/or $\alpha$ are quenched; see  Eq.~\eqref{eq:Hsbm}].
Particularly interesting is the quench that invokes a transition between the coherent 
and incoherent regimes of the dynamics. In this case the difference in the time 
dependencies is particularly clear. The quench dynamics constitutes the main focus of our work.
We investigate quenches of the system-bath coupling as well as those of the Zeeman splitting 
of the up and down spin. In addition we leave the framework of the SBM and identify 
a part of the SBM physics in a bias voltage driven IRLM with two leads.
As the standard relaxation dynamics forms the basis for the understanding 
of the quench ones a detailed study of the former is presented first. In particular, we note
that an understanding of the crossover between coherent and incoherent dynamics requires
the knowledge of the time dynamics on intermediate time scales, which is accessible by
our methods, numerically as well as analytically. Parts of our results for the quench dynamics 
were earlier published in a letter-style publication.\cite{Kennes13b}

At $U=0$, that is $\alpha=1/2$,  the fermionic quantum dot model \eqref{eq:Hirlm} becomes 
noninteracting and can be solved exactly (Toulouse limit). For finite interactions we use
the RTRG and FRG method. The RTRG was specifically developed as an analytical tool to study the 
nonequilibrium physics of small interacting quantum systems coupled to 
noninteracting reservoirs in the weak coupling 
limit.\cite{Schoeller09,Schoeller09_2}  
In particular, the method has the advantage that it can address each individual term
contributing to the time evolution and it can provide analytical insight as to what the generic form of the
time evolution looks like.\cite{Pletyukhov10,Kashuba13}
The method has already been applied to the stationary properties of the IRLM and to the
time evolution for $\epsilon\gg\Gamma_0$.\cite{Andergassen11} 
In this paper, we will discuss the solution of the RG equations and the time dynamics 
for the more difficult resonant case $\epsilon=0$. Moreover, we will generalize the 
RTRG method to the case of arbitrary time-dependent Hamiltonians and will discuss in
detail the quench dynamics for an abrupt change of some system parameter.
The second method applied in this paper is the FRG, which is a very flexible RG method and 
can be used for open as well as closed quantum many-body systems of different 
dimensionality.\cite{Metzner12} The practical implementation for the time evolution 
in the model at hand has been developed in Refs.~\onlinecite{Kennes12,Kennes13a}. 
We discuss in detail how the combined
use of the two methods provides controlled analytical as well as numerical 
access to the standard relaxation and the quench dynamics of the IRLM for 
small $|U|$ that is close to the Toulouse point. 
The time evolution of the SBM in other parameter regimes and for other nonequilibrium setups 
was investigated in 
Refs.~\onlinecite{Grifoni98,Keil01,DiVincenzo05,Hackl08,Alvermann09,Orth10b,Orth13,Kast13a,Kast13b} 
using a variety of methods. 

The paper is structured as follows. In the next section, we introduce the RTRG method 
essentially without referring to a specific model and explain in detail how it can be extended to 
time-dependent Hamiltonians, in particular to study the quench dynamics. 
In Sect.~\ref{sec:frg}, we discuss the basic steps to obtain FRG flow 
equations. Next in Sect.~\ref{sec:appli} we apply both 
RG approaches to the IRLM and study the dynamics within the relaxation and 
quench protocols. This section contains the RG flow equations to be solved as 
well as our analytical and numerical results. We conclude with a brief summary 
in Sect.~\ref{sec:conclu}. The Appendixes contain technical details on the 
generalization of the RTRG method to time-dependent Hamiltonians and quantum 
quenches together with the flow equations and their solutions for the IRLM.

\section{RTRG method}
\label{sec:rtrg}

\subsection{Basic concepts}

The main goal of the RTRG method is to compute the reduced
density matrix of a small interacting  quantum system coupled to several reservoirs by integrating
out the degrees of freedom of the noninteracting bath. From this, the observables 
of interest can be extracted. The reservoirs are initially 
decoupled from the interacting system and are held in thermodynamic equilibrium.
Wick's theorem, which is applicable for the ladder operators of the bath, allows us to 
write a diagrammatic series for an effective Liouvillian which determines the time
evolution of the reduced density matrix of the local quantum system.\cite{Schoeller09}
For the SBM (Ref.~\onlinecite{Kashuba13}) at small $\alpha$ and the IRLM\cite{Andergassen11} at small $|U|$, that 
is the SBM for $\alpha$ close to $1/2$, the perturbative series contains logarithmic divergencies,  
which can be regularized either by cutting off the Matsubara frequencies characterizing the
poles of the Bose/Fermi-functions of the reservoirs\cite{Schoeller09} or by using the Laplace
variable $E$ as a flow parameter. The latter is exploited in the recently 
developed E-RTRG method.\cite{Pletyukhov12} In both cases one obtains RG flow 
equations, the solution of which gives the effective Liouvillian.

Here in the main text we qualitatively summarize the ideas of the RTRG method and its extension to 
time-dependent Hamiltonians and quenches; for technical details and the precise diagrammatic rules
we refer the interested reader to Appendices~\ref{sec:app:diagrams} and \ref{sec:app:quench_dynamics}. 

The time dependence of the reduced density matrix $\rho(t)$ of the local 
quantum system is determined by the von Neumann equation. Due to the coupling to the bath, the system 
acquires a memory which is stored in the form of excitations in the leads.
This results in a time-dependent effective Liouvillian $L(t,t')$, defined for times $t>t'$, which acts 
in Liouvillian space of the local system and determines the dynamics of the reduced density matrix
via the effective von Neumann equation
\begin{equation}
i\dot{\rho}(t) = \int_{t_0}^{t} L(t,t') \rho(t') dt',
\label{eq:general}
\end{equation}
where the coupling of the system to the bath at time $t=t_0$ is implied (later on we will set 
$t_0=0$ for convenience). The effective Liouvillian can be decomposed as
\begin{equation}
\label{eq:sigma}
L(t,t')\,=\,L_S(t)\,\delta(t-t'-0^+)\,+\,\Sigma(t,t') ,
\end{equation}
where $L_S(t)$ is the bare Liouvillian of the isolated local quantum system and $\Sigma(t,t')$
is the dissipative part of the kernel emerging from the coupling to the reservoirs. Defining
for $t>t'$ a propagator $\Pi(t,t')$ relating the reduced density matrix at time $t$ to the
one at $t'$ where system and bath are assumed to be decoupled, we can write the solution of the 
kinetic equation formally as
\begin{equation}
\label{eq:solution_kinetic_equation}
\rho(t)\,=\,\Pi(t,t_0)\,\rho(t_0) ,
\end{equation}
where the relation between $\Pi(t,t')$ and $L(t,t')$ is given by
\begin{equation}
\label{eq:prop_L_relation}
\Pi(t,t') \!=\! \theta(t \!-\! t') \!-\! i \!\!\iint\!\! \theta (t \!-\! t_1)
L(t_1,t_1^\prime)\Pi(t_1^\prime,t') dt_1 dt_1^\prime  .
\end{equation}
Here, all functions depending on two time arguments are retarded 
ones, i.e., are defined as zero for negative time differences.
The last equation expresses the physical property that $-i L(t,t')$ contains
the sum of all correlated processes between $t'$ and $t$.
The rules for the classification of all these processes by a diagrammatic expansion
in the system-bath coupling are provided in Appendix~\ref{sec:app:diagrams}.

For a time-independent Hamiltonian, the effective Liouvillian is a function of the time 
difference only. The resulting von Neumann equation
\begin{equation}
i\dot{\rho}(t) = \int_{t_0}^{t} L(t-t') \rho(t') dt',
\label{eq:beforequench}
\end{equation}
can be formally solved by means of a half-sided Fourier transform (or Laplace transform with the rotated Laplace variable $z\rightarrow -iE$)
\begin{align}
\nonumber
\rho(t)\,&=\,\Pi(t-t_0)\,\rho(t_0) ,\\
\label{eq:relaxPi}
\Pi(t)\,&=\,\frac{1}{2\pi}\,\int_{-\infty+i0^+}^{+\infty+i0^+} e^{-iEt}\,\Pi(E) dE ,
\end{align}
where 
\begin{equation}
\label{eq:prop_laplace}
\Pi(E)\,=\,\frac{i}{E-L(E)}
\end{equation}
is the propagator containing the effective Liouvillian 
\mbox{$L(E)=\int_{0}^{\infty}L(t)e^{iEt}dt$} in Laplace representation. 
The analytical continuation of the propagator can be performed from the real axis 
into the complex plane. As a response function with $\Pi(t<0)=0$, $\Pi(E)$ is an analytical function 
in the upper half-plane. Thus, the Laplace integral transforms into an integral over 
the contour $\mathcal C$ which goes around the nonanalyticities of the propagator 
in the lower half-plane. The general structure of those has been investigated 
in Refs.~\onlinecite{Pletyukhov10,Kashuba13}.
In short, the real and imaginary parts of the pole positions $z_i=\Omega_i-i\Gamma_i$ determine the 
oscillation frequencies $\Omega_i$ and decay rates $\Gamma_i$ of an exponential decay. 
Poles with a finite real part occur in pairs $z_i^\pm=\pm\Omega_i-i\Gamma_i$.
One pole is always located at $E=0$ corresponding to the stationary state 
$L(i0^{+})\rho_{\rm stat}=0$. Additionally, branch cuts occur from the nonanalytic function $L(E)$, 
which contains typically logarithms or power laws arising from logarithmic divergencies in the 
perturbative series. For the time evolution they lead to pre-exponential functions
which, in the long-time limit $T_K t\gg 1$, are typically proportional to
$(T_K t)^{-k}F(g\ln[T_K t])$ with some model-dependent integer $k=0,1,\dots$ and a
slowly varying logarithmic function $F(g\ln[T_K t])$, where $T_K$ is a typical low-energy scale (analog of the Kondo temperature) and $g$ is a
dimensionless coupling constant. 

The non-Markovian contribution to the dynamics is encrypted in the $E$ dependence of 
the effective Liouvillian $L(E)$. $L(E)$ is computed by means of the RTRG 
approach.\cite{Schoeller09,Pletyukhov12} As proposed in Ref.~\onlinecite{Pletyukhov12}, it
is convenient to decompose the Liouvillian as
\begin{equation}
\label{eq:L_decomposition}
L(E)\,=\,L_\Delta(E)\,+\,E\,L'(E) ,
\end{equation}
where $L_\Delta(E)$ and $L'(E)$ are slowly varying logarithmic functions. The idea of the
E-RTRG method,\cite{Pletyukhov12,Kashuba13} where the Laplace variable $E$ is used as the flow 
parameter, is to set up equations for the derivatives $\partial_E L_\Delta(E)$ and
$\partial_E L'(E)$, and to resum the series on the right-hand side (r.h.s.) of 
the differential equation in terms of effective propagators and effective 
vertices. Closing the set of differential equations
by deriving corresponding differential equations for the effective vertices, one obtains
universal RG equations in the infinite band-width limit $\omega_c\rightarrow\infty$ which are free of 
logarithmic divergencies. Provided that the effective vertices stay small (the so-called 
weak-coupling limit), the RG equations can be solved perturbatively in the renormalized 
couplings along an arbitrary path in the complex plane, providing analytical access to an individual 
study of all singularities and branch cuts in the lower half of the complex plane.

Once the parts $L_\Delta(E)$ and $L'(E)$ of the effective Liouvillian are known, the
propagator Eq.~(\ref{eq:prop_laplace}) can be written as
\begin{align}
R(E)&=-i\Pi(E)
\nonumber\\
&=\frac{1}{E-\tilde{L}_\Delta(E)}Z'(E)=\tilde{R}_\Delta(E)Z'(E),
\label{eq:tilde_prop}
\end{align}
where we have defined
\begin{equation}
\label{eq:Zprime_tildeL_def}
Z'\,=\,\frac{1}{1\,-\,L'(E)} ,\quad
\tilde{L}_\Delta(E)\,=\,Z'(E)\,L_\Delta(E)\,,
\end{equation}
and 
\begin{equation}
\label{eq:tilde_resolvent}
\tilde{R}_\Delta(E)\,=\,\frac{1}{E\,-\,\tilde{L}_\Delta(E)} .
\end{equation}
For the special case of the IRLM, the RG equations for $\tilde{L}_\Delta(E)$ and
$L'(E)$ together with the results for the effective vertices have been derived in
Ref.~\onlinecite{Andergassen11} using the Matsubara cutoff scheme of Ref.~\onlinecite{Schoeller09}.
The same equations can be obtained within $E$-RTRG by using the Laplace variable as flow parameter.\cite{Schoeller13}
The results are summarized in Appendix~\ref{sec:app:rtrg_irlm} and
will be used in this work as a starting point to analyze the dynamics within the relaxation and
quench protocols.

\subsection{Extension to quenches}
\label{qRTRG}

In this paper, we extend the RTRG method to quenches, namely, abrupt parameter changes 
in the Hamiltonian. The system is described by the Hamiltonian $H_{\text{tot}}^{i}$ before the time of 
the quench $t_{q}$ and by $H_{\text{tot}}^{f}$ afterwards, while both $H_{\text{tot}}^{i}$ 
as well as $H_{\text{tot}}^{f}$ are assumed
to be \emph{time independent.} In this case, the integral in Eq.~\eqref{eq:general} can be split into 
two parts that describe the memory of the system before and after $t_{q}$. Equation 
\eqref{eq:beforequench} determines the density matrix 
$\rho_i(t)=\rho(t)\theta(t_q-t)$ before the quench, 
where the Liouvillian $L_{i}$, calculated 
using $H_{\text{tot}}^{i}$, is taken. In the equation describing the dynamics of the 
density matrix $\rho_f(t)=\rho(t)\theta(t-t_q)$ after the quench there is an 
additional term containing the memory of the systems dynamics before the quench:
\begin{eqnarray}
i\dot{\rho}_f(t) &=& \int_{t_{q}}^{t} L_{f}(t-t') \rho_f(t') dt'
\nonumber\\&+&
\int_{t_0}^{t_{q}}\Sigma_{fi}(t,t')\rho_i(t')dt' , 
\label{eq:afterquench}
\end{eqnarray}
where the Liouvillian $L_{f}$ is calculated using the Hamiltonian $H_{\text{tot}}^{f}$ after the quench, and the 
kernel $\Sigma_{fi}(t,t')$ describes the system's memory of processes that took 
place before the quench. Therefore, $\Sigma_{fi}(t,t')$ is only defined for $t'<t_q<t$.

Most conveniently, the solution of 
Eq.~\eqref{eq:afterquench} can be written in terms of the propagator 
$\Pi_{fi}(t,t')=\theta(t-t_q)\Pi(t,t')\theta(t_q-t')$,
which connects the density matrix after the quench with the initial density matrix before the
quench:
\begin{equation}
\label{eq:rho_Pi}
\rho_f(t) \,=\, \Pi_{fi}(t,t_0)\,\rho(t_0) .
\end{equation}
Using Eq.~(\ref{eq:prop_L_relation}), this propagator can be split into two parts
\begin{align}
\nonumber
\Pi_{fi}(t,t')\,&=\,\Pi_f(t,t_q)\,\Pi_i(t_q,t')\\
\label{eq:Pi_split}
&\hspace{-1.5cm}-\,i\,\iint
\,\Pi_f(t,t_1)\,\Sigma_{fi}(t_1,t_1^\prime)\,\Pi_i(t_1^\prime,t')  dt_1 dt_1^\prime ,
\end{align}
where $\Pi_{f/i}(t,t')=\Pi_{f/i}(t-t')$ are only defined for $t>t'$ and are the propagators 
of a system time evolved for all times with $H_{\text{tot}}^{i/f}$ taken from 
Eqs.~(\ref{eq:relaxPi}) and (\ref{eq:prop_laplace}) 
with Liouvillian $L_{i/f}$. Inserting Eq.~(\ref{eq:Pi_split}) in~(\ref{eq:rho_Pi}) and
using $\Pi_i(t_q-t_0)\rho(t_0)=\rho(t_q)$ we obtain
\begin{align}
\nonumber
\rho_f(t)\,&=\,\Pi_f(t-t_q)\,\rho(t_q)\\
\label{eq:time_evolution_split}
&\hspace{-1cm}-i\iint
\Pi_f(t-t_1)\Sigma_{fi}(t_1,t_1^\prime)\Pi_i(t_1^\prime-t_0)\rho(t_0)  dt_1 dt_1^\prime .
\end{align}
The first term describes the dynamics without any memory to times smaller than the quench time, i.e. the density matrix 
$\rho(t_q)$ at the quench time is used as initial condition for the time evolution after 
the quench as if the local system and the reservoirs were decoupled up to time $t_q$.
The second term describes the memory part where the quench time is inside the memory
kernel $\Sigma_{fi}$. 

\begin{figure}
\centering
\includegraphics[scale=.5]{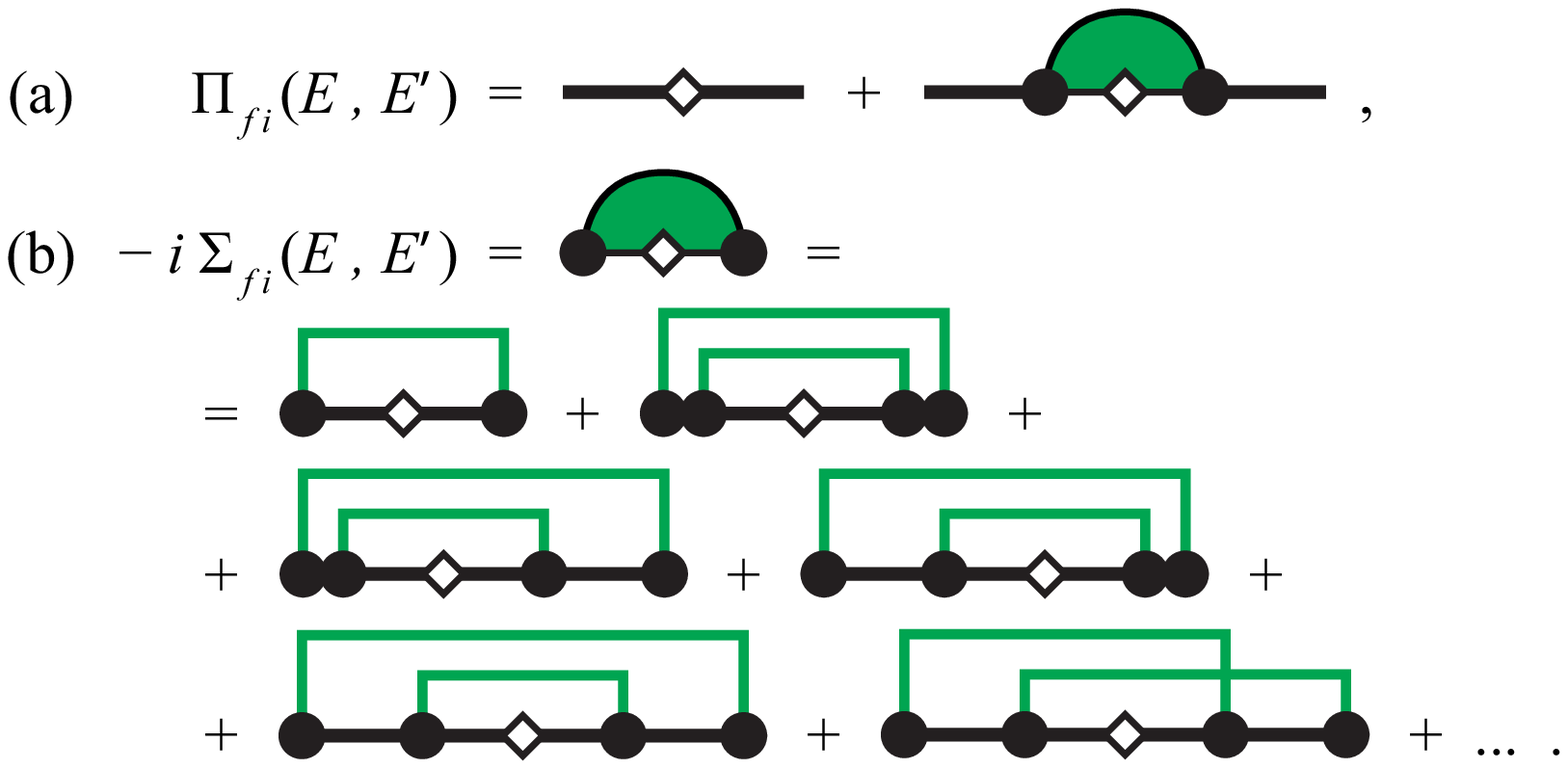}
\caption{(Color online) 
(a) Diagrammatic view of a quench propagator corresponding to Eq.~\eqref{eq:Pi_split} (time representation) or 
Eq.~\eqref{eq:pieedyson} (Laplace representation).  
(b) Diagrammatic series for a quench kernel; the first diagrams is 
expressed in Eq.~\eqref{eq:sigma12}.
The thick horizontal lines are propagators of a local system; diamonds are quench vertices; single circles depict tunneling vertices, while doubled circles represent an interaction vertex; the green lines connecting 
circles correspond to reservoir contractions.}
\label{fig:diagrams}
\end{figure}

Defining the double Laplace transform
\begin{equation}
\label{eq:quenchPi}
\Pi_{fi}(E,E')=\iint\!\! e^{iE(t-t_q)-iE'(t'-t_q)}\,\Pi_{fi}(t,t')  dt dt' ,
\end{equation}
and analogous for $\Sigma_{fi}(E,E')$, Eq.~(\ref{eq:Pi_split}) can be written in Laplace space as
\begin{equation}
\Pi_{fi}(E,E') = \Pi_{f}(E)\bigl[ 1-i\Sigma_{fi}(E,E')\bigr]\Pi_{i}(E').
\label{eq:pieedyson}
\end{equation}
Using the inverse Laplace transform of the second term on the r.h.s. of this
equation and substituting it for the second part of the propagator Eq.~\eqref{eq:Pi_split},
we can write the time evolution Eq.~\eqref{eq:time_evolution_split} as
\begin{align}
\nonumber
\rho_f(t)\,&=\,\frac{1}{2\pi}\,\int e^{-iE(t-t_q)}\,\Pi_f(E)\,\rho(t_q) dE\\
\nonumber
& \hspace{0.5cm} -i \iint e^{-iE(t-t_q)-iE'(t_q-t_0)}  \\
\label{eq:time_evolution_quench}
& \hspace{0.5cm} \times \Pi_f(E)\,\Sigma_{fi}(E,E')\,\Pi_i(E')\,\rho(t_0) \frac{dE dE'}{(2\pi)^2} .
\end{align}

The two terms of Eq.~\eqref{eq:pieedyson} and the diagrammatic expansion of the memory kernel 
in terms of the coupling vertices are illustrated in Laplace space in 
Fig.~\ref{fig:diagrams}(a) and Fig.~\ref{fig:diagrams}(b). Here, the quench vertex is indicated as a diamond and is 
implemented in the diagrammatics as a unit matrix vertex:
\begin{equation*}
\text{quench vertex: } \hat{1}\, = \, \raisebox{-1ex}{\includegraphics[scale=.5]{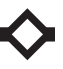}} . 
\end{equation*}
As implied by Eq.~\eqref{eq:pieedyson} all operators on the left of the quench vertex depend on the 
Laplace variable $E$ and have to be taken w.r.t. $H_{\text{tot}}^{f}$ , while the ones on the right depend on 
$E'$ and are taken w.r.t. $H_{\text{tot}}^{i}$. 
The diagrammatic rules are explained in detail in Appendix~\ref{sec:app:quench_dynamics},
some examples of diagrams are shown in Fig.~\ref{fig:diagrams}(b) for the IRLM. The diagrams consist of 
vertices with one or two reservoir lines, corresponding to tunneling and Coulomb interaction 
vertices, respectively. The vertices are connected by propagators $R=-i\Pi$ (horizontal black lines) and 
reservoir contractions (green lines). In each diagram the quench vertex has to be 
inserted in every bare propagator.
Analogous to Ref.~\onlinecite{Pletyukhov12}, we resum all diagrams with excitations 
living only before and after the quench (i.e.~contractions which do not cross over the quench) 
such that full effective propagators and vertices occur to the left and right of the quench vertex. 
As a consequence only diagrams with contractions crossing over the quench vertex have to be considered. 
Sorting diagrams by the number of contractions, i.e. the number 
of excitations gone through the quench, the series starts as shown in Fig.~\ref{fig:diagrams}(b).
In this we present all diagrams with one or two contractions. 
Using (\ref{eq:sigma_fi_diagrams_effective}) and (\ref{eq:contraction_explicit}) and considering the
IRLM with only one reservoir, chemical potential $\mu=0$, and zero temperature, the first diagram 
is explicitly given by (for the notation see Appendices~\ref{sec:app:diagrams} and \ref{sec:app:quench_dynamics})
\begin{align}
\nonumber
-i\,\Sigma_{fi}(E,E')\,&=\,\int p'f(p'\bar{\omega})\, G^f_{1}(E) \\
\nonumber
&\hspace{-1.5cm}
\times R_{f}(E+\bar{\omega})\,R_{i}(E'+\bar{\omega})\,
(G_i)_{\bar{1}}^{p'}(E'+\bar{\omega}) d\bar{\omega} ,
\end{align}
where $f(\omega)=\theta(-\omega)$ is the Fermi function at $T=0$,
$\bar{\omega}=\eta\omega$, $G^{f/i}_1\equiv\sum_p (G_{f/i})^p_1$ and 
$R_{f/i}(E)=1/[E-L_{f/i}(E)]$. No frequency cutoff is needed since the integral
converges. Neglecting the frequency dependence of the second 
vertex (giving rise to higher order terms) and closing the integration contour in the 
upper half of the complex plane we see that only the nonanalytic part of 
$p'f(p'\bar{\omega})=-\frac{1}{2}\text{sign}(\bar{\omega})+p'/2$ contributes to the integral. 
The integral around the branch cut of the $\text{sign}$-function at $\bar{\omega}=i\Lambda$,
$0<\Lambda<\infty$, leads to  
\begin{align}
\nonumber
\Sigma_{fi}(E,E')\,&=\,\int_0^\infty
 (G_f)_1(E)\,R_{f}(E+i\Lambda) \\
\label{eq:sigma12}
&\hspace{-1.5cm}
\times R_{i}(E'+i\Lambda)\,(G_i)_{\bar{1}}(E') d\Lambda
\end{align}
This result will be used in Sec.~\ref{sec:quench_protocols} to analyze the quench dynamics
via Eq.~\eqref{eq:time_evolution_quench}. All other diagrams of Fig.~\ref{fig:diagrams}(b) are
unimportant. The second diagram is divergent and has to be treated by the E-RTRG method by
considering its derivative w.r.t. $E$.\cite{Pletyukhov12} However, it contributes only to 
the dynamics of the off-diagonal terms of the reduced density matrix, which we do not consider here. 
The integrals over the frequencies in all other diagrams do not diverge and therefore computing those does 
not require a RG procedure. They can be directly evaluated but lead to higher order terms.

\section{FRG method}
\label{sec:frg}

The FRG is a flexible method which allows to tackle a variety of open
as well as closed quantum many-body models within the same framework.\cite{Metzner12} 
In this approach one aims at the one-particle irreducible vertex functions from which 
observables can be computed. 
The formulation of the FRG approach in Matsubara space provides direct access to the 
equilibrium properties of microscopic many-body models. It was used for studying 
quantum dots, quantum wires and two-dimensional lattice models.\cite{Metzner12} 
The FRG was extended to Keldysh space and steady-state nonequilibrium 
properties\cite{Gezzi07,Jakobs07,Jakobs10,Karrasch10} as well as nonequilibrium 
time evolution\cite{Kennes12,Kennes12b,Kennes13a} was studied. To keep the presentation simple 
we focus on the Keldysh formalism (used in this paper) when presenting explicit formulas from now on. 

\begin{figure}[t]
\centering
\includegraphics[width=.7\columnwidth,clip]{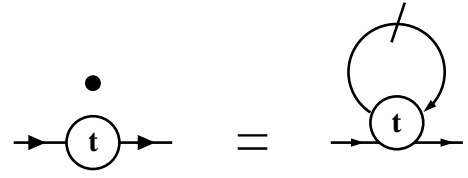}
\caption{Diagrammatic representation of the self-energy flow equation. The dot indicates 
the derivative w.r.t.~$ \Lambda $ and the slanted line is the single-scale propagator 
$S^\Lambda$.}
\label{fig:Flowdia}
\end{figure}

In a first step of the derivation of flow equations one introduces a cutoff 
$\Lambda$ to the free propagator
\begin{equation}
G_0\rightarrow G^{\Lambda}_0,\;\;\;G^{\Lambda=\infty}_0=0,\;\;\;G^{\Lambda=0}_0=G_0 ,
\end{equation}
which is a 2$\times$2-matrix in Keldysh contour space. 
During the RG flow this cutoff is removed and the problem of interest is restored. One 
takes the derivative of the generating functional $\Gamma^\Lambda(\{\bar\phi\},\{\phi\})$ 
for the irreducible vertex functions with respect
to $\Lambda$ 
\begin{equation}
\begin{split}
\dot \Gamma^\Lambda(\{\bar\phi\},\{\phi\})=&{\text{Tr}}\; \left[G_0^\Lambda\partial_\Lambda
\left[G_0^\Lambda\right]^{-1}\right]\\&\mp i {\text{Tr}}\; \left[\partial_\Lambda\left[G_0^\Lambda\right]^{-1}
\frac{\delta^2\mathcal{W}^{c,\Lambda}}{\delta\bar\eta^\Lambda\delta\eta^\Lambda}\right] 
\end{split}
\end{equation}
and expands with respect to the source fields $\bar\phi,\phi$. 
Aside from the usual dependence on the quantum numbers and times, the latter carry 
an additional index for the upper or lower Keldysh contour.
Here, the upper (lower) sign is for bosons (fermions) and we defined the 
generating functional of the connected Green's functions by
\begin{equation}
\begin{split}
\mathcal{W}^{c,\Lambda}(\{\bar\eta\},\{\eta\})=\ln\bigg[\frac{1}{\mathcal{Z}_0}\int\mathcal{D}\bar\psi\psi\exp\{S_0^\Lambda-iS_{\text{int}}\\-(\bar \psi,\eta)-(\psi,\bar\eta)\}\bigg]
\end{split}
\end{equation}
with the noninteracting partition function $\mathcal{Z}_0$ and the noninteracting  $S_0^\Lambda$
and interacting $S_{\text{int}}$ part of the action defining the quantum many body problem. 
This leads to an infinite hierarchy of flow equations for the vertex functions, with the 
first three given by
\begin{equation}
\begin{split}
\partial_\Lambda\gamma_0^\Lambda &= {\text{Tr}}\;\left[G_0^\Lambda\partial_\Lambda [G_0^\Lambda]^{-1}\right] - 
{\text{Tr}}\;\left[ G^\Lambda\partial_\Lambda [G_0^\Lambda]^{-1}\right]~\\
\partial_\Lambda\gamma_1^\Lambda(1';1) &= \sum_{2 2'}S_{2 2'}^\Lambda
\gamma_2^\Lambda(1'2';1 2)\\
\partial_\Lambda\gamma_2^\Lambda(1'2';1 2) &=  
\sum_{3 3'}  S^\Lambda_{3 3'} 
\gamma_3^\Lambda(1 '2' 3' ;1 2 3) \\
&-\sum_{3 3' 4 4'}S^\Lambda_{3 3'}\gamma_2^\Lambda(3' 4;1 2)
G^\Lambda_{4' 4} \gamma_2^\Lambda(1' 2';4' 3)\\
&-\Bigg[\sum_{3 3' 4 4'}S^\Lambda_{3 3'}
\gamma_2^\Lambda(1' 3';1 4)
G^\Lambda_{ 4 4'}\gamma_2^\Lambda(2' 4';2 3)\\
& \hspace*{1cm} - ( 1' \leftrightarrow 2') - ( 1 \leftrightarrow  2) \\
& \hspace*{1cm}+ ( 1'\leftrightarrow  2', 1 \leftrightarrow  2) \Bigg] ,\\
\end{split}
\end{equation}
where we have used
\begin{equation}
S^\Lambda_{ 1' 1} = S^\Lambda(1' , 1) = \sum_{2 2' } 
G_{1' 2}^\Lambda\left[\partial_\Lambda [G_0^\Lambda]^{-1}\right]_{2 2'}G_{2' 1}^\Lambda
\end{equation} 
and an appropriate multi-index $1,1',\ldots$ denoting the quantum numbers, time, 
and the contour label. In practical applications to time evolution, one is restricted to the lowest-order (in the 
two-particle interaction $U$) truncation scheme of the resulting coupled differential 
equations for the vertex functions (see Fig.~\ref{fig:Flowdia}).
This results in differential equations for the Keldysh 
components of the self-energy which are controlled to leading order in the interaction but due
to the RG resummation go beyond plain perturbation theory; e.g.~in a variety of applications 
one obtains power laws with interaction dependent exponents.\cite{Metzner12} The flow equations
can be solved numerically with minor computational effort which allows to study the 
entire parameter space of a given model with high efficiency. In limiting cases one often 
even succeeds in gaining analytical insights from these flow equations.  Within the 
time-dependent FRG approach one can straightforwardly treat quenches as all parameters of the
Hamiltonian can carry an explicit time dependence. This was discussed in detail in 
Ref.~\onlinecite{Kennes12b}.

\section{Application to dissipative dynamics}
\label{sec:appli}

In this section we apply the RG methods introduced above to the 
IRLM and describe its dynamics. We restrict ourselves to the small-$|U|$ regime 
corresponding to the SBM close to its coherent-to-incoherent
transition. Ultimately, we are interested in the quench dynamics but first present 
a comprehensive study of the relaxation protocol. It presents the basis for the
understanding of the quench protocol.  

\subsection{Relaxation protocol}
\label{relpro}

\subsubsection{The RTRG approach}

The general Eqs.~(\ref{eq:relaxPi}) and (\ref{eq:prop_laplace}) can be used to describe the 
behavior of the system between the initial time $t_0=0$ of coupling 
the reservoir to the dot and the time of the quench $t_{q}$. Using the form
Eq.~(\ref{eq:tilde_prop}) for the propagator, we obtain
\begin{equation}
\label{eq:rho(t)}
\rho(t)\,=\,\frac{i}{2\pi}
\int_{\mathcal{C}}e^{-iEt}\,\tilde{R}_\Delta(E)\,Z'(E)\,\rho(0)  dE,
\end{equation}
where $\mathcal{C}$ is a contour enclosing clockwise the lower half of the complex plane 
including the real axis.
 
For the particle-hole symmetric case $\epsilon=0$, the $4\times 4$ matrices $\tilde{R}_\Delta(E)$ 
and $Z'(E)$ are provided in Eqs.~(\ref{eq:Zprime_matrix}) and (\ref{eq:tilde_R_matrix}).
Since we are only interested in the relaxation of the diagonal matrix elements
$\rho_{00}(t)$ and $\rho_{11}(t)$ of the density matrix, we obtain
\begin{align}
\nonumber
\begin{pmatrix}\rho_{00}(t)\\ \rho_{11}(t)\\\end{pmatrix}
&=\frac{i}{2\pi}\int_{\mathcal{C}}\,e^{-iEt}\\
\label{eq:rho_diagonal(t)_1}
&\times \left(\frac{\tau_+}{E}\,+\,\frac{\tau_-}{E+i\Gamma_1(E)}\right)
\begin{pmatrix}\rho_{00}(0)\\ \rho_{11}(0)\\\end{pmatrix}  dE ,
\end{align}
with $\tau_\pm=\frac{1}{2}(1\pm\sigma_x)$. Using 
$\frac{i}{2\pi}\int_{\mathcal{C}}e^{-iEt}\frac{1}{E}=1$ and $\rho_{00}(0)+\rho_{11}(0)=1$, we find
\begin{align}
\nonumber
\begin{pmatrix}\rho_{00}(t)\\ \rho_{11}(t)\\\end{pmatrix}
\,&=\,\frac{1}{2}\,\begin{pmatrix}1 \\ 1 \\\end{pmatrix}\\
\label{eq:rho_diagonal(t)_2}
& \hspace{-1cm}
+\,\frac{i}{2\pi}\int_{\mathcal{C}}
\frac{e^{-iEt}}{E+i\Gamma_1(E)}\frac{1}{2}
\begin{pmatrix}-1\\ 1\\\end{pmatrix}\,\langle \sigma_z \rangle(0) dE ,
\end{align}
with $\langle\sigma_z\rangle(0)=\rho_{11}(0)-\rho_{00}(0)$. This gives
\begin{equation}
\label{eq:sigma_z}
\langle \sigma_z \rangle(t)\,=\,P(t)\,\langle\sigma_z\rangle(0) ,
\end{equation}
where 
\begin{equation}
P(t) = \frac{i}{2\pi}\int_{\mathcal C} \frac{e^{-iEt}}{E+i\Gamma_1(E)} dE
\label{eq:Prelax}
\end{equation}
is defined as the solution for $\langle\sigma_z\rangle(t)$ with the initial condition 
$\langle\sigma_z\rangle(0)=1$.

To determine $P(t)$ we need the function $\Gamma_1(E)$, which follows from
the RG equations (\ref{eq:RG_Gamma_1}) and (\ref{eq:RG_Gamma_2}):
\begin{equation}
\partial_{E}\Gamma_{1/2}(E) \!=\! -g R_{2/1}(E)\Gamma_{1}(E)\,,
\label{eq:rgflow}
\end{equation}
where $g=2U-U^{2}$ and 
\begin{equation}
R_{1}(E) = \frac{1}{E+i\Gamma_{1}(E)},\,\,\,R_{2}(E) = \frac{1}{E+i\Gamma_{2}(E)/2} ,
\label{eq:resolvent}
\end{equation}
with the initial conditions \mbox{$\Gamma_{1/2}(E=i\omega_c)=\Gamma_{0}$}. For
high energies $|E|\gg |\Gamma_{1/2}(E)|$, the solution of the RG equations is
\begin{equation}
\label{eq:solution_high_energies}
\Gamma_{1/2}(E)\,=\,\Gamma_0\,\left(\frac{\omega_c}{-iE}\right)^g\,=\,
T_K\,\left(\frac{T_K}{-iE}\right)^g ,
\end{equation}
where 
\begin{equation}
\label{eq:T_K}
T_K\,=\,\Gamma_0\,\left(\frac{\omega_c}{ T_K}\right)^g\,=\,
\Gamma_0\,\left(\frac{\omega_c}{\Gamma_0}\right)^{g/(1+g)}
\end{equation}
is the typical low-energy scale which is kept fixed in the scaling limit
$\omega_c\rightarrow\infty$ and $g,\Gamma_0\rightarrow 0$. In this limit, the 
RG equations (\ref{eq:rgflow}) cover systematically all orders
of $g\ln(\omega_c/E)$ and the exponent of the power law in Eq.~(\ref{eq:solution_high_energies})
is controlled up to ${\mathcal O}(U^2)$. The terms neglected on the r.h.s. of the RG equation
for $\Gamma_1(E)$ contain terms $\sim U^3(\Gamma_1/E)$ of higher orders in $U$ (which
can be neglected for small $U$) and terms $\sim U(\Gamma_{1/2}/E)^2$ of higher order in
$\Gamma_{1/2}/E$. Integrating the latter from $E=i\omega_c$ down to $E\sim i\Gamma_{1/2}$ gives a
correction to $\Gamma_1$ of order ${\mathcal O}(U\Gamma_{1/2})\sim 
{\mathcal O}(U T_K [T_K/\Gamma_{1/2}]^g)$. 
This is a small
correction to Eq.~(\ref{eq:solution_high_energies}) of relative order $U$, which 
changes only the prefactor of the Kondo temperature $T_K$ but does not influence 
the power law exponent. The FRG approach described above can cover such corrections 
up to ${\mathcal O}(U)$ since all orders in the tunneling are fully taken into account in each
order of the Coulomb interaction.

We note, however, that neglecting higher orders in $\Gamma_{1/2}/E$ might imply a
change of the power-law exponent of the time evolution in the regime of exponentially
large times $T_K t\gg 1$ {\it and} $|g\ln(T_K t)|\gtrapprox 1$. In this regime one needs a 
solution of the RG equations for $E$ exponentially close to a singularity $z^*$ of the
resolvents $R_n(E)$ and it may happen that
neglected terms $\sim g(\Gamma_{1/2})^2R_1(E)R_2(E)$ are of order $g|\Gamma_{1/2}/(E-z^*)|$
since one resolvent is of order $1/|z^*|\sim 1/|\Gamma_{1/2}|$
while the other one has a pole at $E=z^*$ and is of order $1/|E-z^*|$.
Such terms lead to additional logarithmic contributions close to the singularities of
the resolvent $R_1(E)$ which might result in logarithmic corrections for pre-exponential functions
relevant for exponentially large times. Recently, it has been
shown\cite{Kashuba13} that such behavior can occur for the SBM at small coupling $\alpha$
and it needs to be analyzed whether similar effects may be realized in the IRLM \cite{Schoeller13}
at small $g$. 
The most interesting physics occurs when different terms in the dynamics compete. 
For the relaxation as well as the quench protocols this happens at intermediate 
($T_K t \sim 1$) to long ($T_K t\gtrapprox 1$) times while 
at exponentially large ones $|g\ln(T_K t)|\gtrapprox 1$ one term dominates (see below). 
For the former we can safely ignore higher-order terms in $\Gamma_{1/2}/E$. We note that for 
completeness and to make contact to existing results we also analyze our approximate 
RG equations for  exponentially large times keeping in mind that the corresponding result might
not present the final answer.

\subsubsection{The FRG approach}

\begin{figure}
\centering
\includegraphics[width=.7\columnwidth,clip]{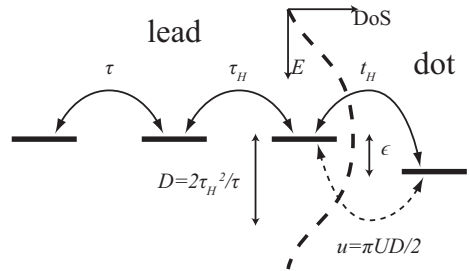}
\caption{Quantum dot model considered within FRG.}
\label{fig:QDfrg}
\end{figure}

In the FRG approach to the IRLM [Eq.~(\ref{eq:Hirlm})], we consider a semi-infinite 
tight-binding chain, where the first site is tunnel coupled to a single level almost resonant to the 
Fermi level in the chain (see Fig.~\ref{fig:QDfrg}). 
A fermion occupying the first site of the chain interacts with the resonant level by a 
density-density--type interaction of strength $u$. The first lead site and the 
single level define our quantum dot region. Initially the two sites of the quantum dot region are empty. With the rest of the chain we proceed as follows.
We achieve a structureless reservoir by choosing the hopping $\tau$ between the corresponding sites 
and the hopping $\tau_H$ from the second to the first site  such that 
$\tau\to \infty$, $\tau_H\to \infty$, but $\tau_H^2/\tau$ remains constant.
As long as the hopping $t_H$ between the first site of the chain and the resonant level 
remains small compared to the bandwidth $\omega_c= 2 \tau_H^2/\tau$, we capture the physics 
of the IRLM as defined in Eq.~\eqref{eq:Hirlm}, where the parameters are mapped by choosing 
$\Gamma_0=4t_H^2/\omega_c$ and $U=2u/(\pi \omega_c)$.\cite{Karrasch10,Kennes12}   

The FRG flow equations to be solved numerically were already derived in detail in 
Ref.~\onlinecite{Kennes12}. The main steps and essential equations are summarized 
at this point to keep our presentation self-contained. The nonequilibrium 
problem at hand is tackled by using the Keldysh formalism.\cite{HaugJauho,Rammer} 
First, we determine the retarded (R), advanced (A), and Keldyh (K) components 
of the Green's function of the decoupled and noninteracting dot 
region
\begin{align}
  \label{eq:gRet}
  g^\Ret_{ij}(t,t') &= - i \Theta(t-t')  \left[ e^{-i 
    \hat\epsilon (t-t') }\right]_{ij},
  \\
  \label{eq:gK}
  g^\K_{ij}(t,t') &= -i \left[g^\Ret(t,0) g^\Adv(0,t')\right]_{ij}.
\end{align}
Here and in the following the two sites of the quantum dot region are labeled 
by the single-particle indices $i,j\in \{1,2\}$.
The matrix $\hat\epsilon$ appearing in the argument of the exponential function 
is given by
\begin{equation}
 \hat\epsilon =\begin{pmatrix}
    t_H & \epsilon
    \\
    0 & t_H
  \end{pmatrix}.
\end{equation}
We next treat the influence of the reservoir and the interaction separately in the form of 
self-energies. To obtain the reservoir part, we project the reservoir 
contribution on the noninteracting dot Green's function $g_0$, exploiting 
that the interaction is restricted to the quantum 
dot region. This way, the influence of the reservoir is considered by tracing over 
the particles temporarily (virtually) residing in the reservoir by tunneling from the dot 
and back. This yields a Dyson equation for the reservoir-dressed but noninteracting 
dot Green's function $G_0$:
 \begin{equation}
 G_0=g_0+g_0\Sigma_{\res} G_0,
\label{dydy}
 \end{equation}
where we have left implicit internal matrix multiplications w.r.t.\ the Keldysh and 
single-particle indices and convolutions w.r.t.\ time. If we restrict 
ourselves to a structureless reservoir, the reservoir contribution to 
the self-energy $\Sigma_{\res}$ is given by
\begin{align}
  \label{eq:SigmaRet_wbl}
  \Sigma^\Ret_{\res,ij}(t',t) &=\left[\Sigma^\Adv_{\res,ij}(t',t)\right]^\dagger = - 2\omega_ci \delta(t'-t) \delta_{i,j}\delta_{i,1},
  \\
  \Sigma^\K_{\res,ij}(t',t) &= -\frac{4\omega_c}{\pi} \mathcal{P}\left(\frac{1}{t'-t}\right)\delta_{i,j}\delta_{i,1},
  \label{eq:SigmaK_wbl}
\end{align}
Using the Dyson equation (\ref{dydy}) is an exact reformulation of the noninteracting problem; 
all orders in the tunneling to the reservoirs are kept.
Next, we consider the Keldysh self-energy $\Sigma$ arising due to the two-particle 
interaction and again employ the Dyson equation 
\begin{equation}
 G=G_0+G_0\Sigma G
 \end{equation} 
to obtain the full Green's function $G$.
The cutoff procedure used in the FRG consists of two independent 
auxiliary reservoirs coupled to each of our two dot sites via hybridization 
$\Lambda$.\cite{Jakobs10} Differentiation of the generating functional 
of the one-particle irreducible vertex functions with respect to $\Lambda$ yields the 
above-mentioned infinite hierarchy of flow equations for the vertex functions which is 
still exact. The truncation of this hierarchy is the only approximation within our approach 
required to derive 
a closed set of differential equations which can be integrated numerically.
We use the lowest truncation order, and the resulting flow equations for the 
interaction part of the self-energy are given by
\begin{eqnarray}
\partial_\Lambda \Sigma^\Lambda(1, 1') &= & -i \sum_{2 , 2'} S_{2' 2}^\Lambda
 \bar u_{1  2 1' 2'},
\end{eqnarray}
with a multi-index $ 1=(t,i,p) $ with $ p\in\{-1,1\} $ labeling the Keldysh index and
\begin{equation}
\begin{split}
 S^\Lambda_{1 1'} &= -\sum_{2 , 2'} G_{1 2'}^\Lambda\left[\partial_\Lambda [ G^{\Lambda}_0]^{-1}\right]_{2' 2} G_{2 1'}^\Lambda\\& = \partial^*_\Lambda  G_{1 1'}^\Lambda.
\end{split}
\end{equation}
We introduced the star differential operator $\partial^*_\Lambda$ which acts only on the free 
Green's function $  G^{\Lambda}_0 $, not on $ \Sigma^\Lambda $, in the series expansion $  G^{\Lambda}=  G^{\Lambda}_0+ G^{\Lambda}_0\Sigma^\Lambda  G^{\Lambda}_0+\dots$ . 
Additionally we define
\begin{equation}
\begin{split}
\bar u_{1 2 1' 2'}=&\delta(t_1-t_{1}')\delta(t_1-t_{2})\delta(t_1-t_{2}')\\
&\times \delta_{p_1,p_1'}\delta_{p_2,p_2'}\delta_{p_1,p_2}(p_1) \bar u_{i_1 i_2 i_1' i_2'}(t) 
\end{split}
\end{equation}
with the anti-symmetrized two-particle interaction $\bar  u_{i_1 i_2 i_1' i_2'} $.
As a consequence of the truncation the FRG results are controlled to leading order in the interaction 
strength $U$. In the RTRG the interaction was expressed in terms of the prefactor $g=2U-U^2$ of
the flow equations \eqref{eq:rgflow}, which in the solution of the latter appears as an 
exponent. In the FRG this exponent is only captured to leading order. To prevent the inflation 
of parameters in the following we still use $g$ when discussing FRG results, keeping in mind this 
restriction to the lowest order.    
The present FRG procedure was successfully applied to nonequilibrium transport 
through correlated quantum dots before.\cite{Jakobs10,Karrasch10,Kennes12}
The cutoff-free problem is recovered after integrating from $\Lambda=\infty$, where one can 
give analytic expressions for the vertices, down to  $\Lambda=0$, 
where the problem corresponds to an effective noninteracting one with renormalized time 
dependent single-particle parameters. 

At the point of particle-hole symmetry $\epsilon=0$ for
$t\gg \omega_c^{-1}$ and to leading order in $g$ the only relevant flow equation 
is the one for the hopping amplitude between the resonant level and the first site of the reservoir. 
It reads
\begin{equation}
\partial_\Lambda t_H^\Lambda(t)=\frac{\pi g}{8 \omega_c}i\partial_\Lambda^\ast G^K_{1,l}(t),
\end{equation} 
where $\partial_\Lambda^\ast$ acts on the free part of the full equal time Keldysh component 
of the Green function $G^K_{1,l}(t)$ only.
Its initial condition is $t_H^{\Lambda=\infty}(t) = t_H$ for all $t$.
Employing the projection technique we arrive at
\begin{multline}
\partial_\Lambda t_H^\Lambda(t)=i\frac{\pi g \omega_c}{4}\partial_\Lambda^\ast\int t_H^\Lambda(t')
\left[g^>(t,t')G^<(t',t)\right. \\
\left. -g^<(t,t')G^>(t',t)\right]dt',
\label{eq:flowth1}
\end{multline}
with the structureless reservoir's Green functions $g^{>(<)}$ and the single level's 
Green functions $G^{>(<)}$. 
The greater and lesser functions are related to the retarded, advanced, and Keldysh 
components by a Keldysh rotation.\cite{HaugJauho,Rammer}

Already in this simple truncation scheme the logarithmic terms present 
in lowest-order perturbation theory in $g$ are resumed consistently. One obtains a 
renormalized hopping amplitude featuring a power law with interaction dependent 
exponent.
This exponent is  correct to leading order in the interaction.\cite{Karrasch10}
It was shown that for the time evolution FRG (Ref.~\onlinecite{Kennes12}) leads to terms 
exponentially decaying in time with interaction-dependent decay rates as well 
as power-law corrections $t^{-\kappa}$ with $g$-dependent exponent $\kappa$ which
is consistent with the one found within RTRG.\cite{Andergassen11} 

To obtain analytical results using FRG we replace  $t_H(t')$ by $t_H(t)$ 
in Eq.~\eqref{eq:flowth1}, which is justified to order $g$.
To lowest order in $1/\omega_c$ we then obtain
\begin{equation}
\begin{split}
\frac{\partial_\Lambda t_H^\Lambda(t)}{t_H^\Lambda(t)}=\frac{g}{2}\partial_\Lambda\!\!\left[-i\int^t_{2/\omega_c}\!\!\!\! G^{A,\Lambda}_0(x)\frac{dx}{x}+\frac{\pi}{2}G^{K,\Lambda}_0(t)\right],\label{eq:flowth2}
\end{split}
\end{equation}
with the noninteracting but reservoir dressed advanced Green function 
$G^{A,\Lambda}_0(x)= i\exp (-\Gamma_0 x/2 ) $ as well as 
$i G^K_0(t)= \exp (-\Gamma_0 t ) $.
The high energy cutoff $\tau_H^2/\tau=\omega_c/2$ has to be introduced by hand as 
the higher order terms in $1/\omega_c$ were dropped in Eq.~\eqref{eq:flowth2}.
Integrating Eq.~\eqref{eq:flowth2} we find 
\begin{equation}
t_H(t)=\frac{\sqrt{T_K \omega_c}}{2} \exp\Bigl[-\frac{g}{2}\Bigl({\rm E}_1(\tfrac{\Gamma_0 t}{2})+\frac{i \pi}{2}e^{-\Gamma_0 t} \Bigr)\Bigr] \label{eq:th_ren}
\end{equation}
with ${\rm E}_1(z)=\int_{z}^{\infty}e^{-x}dx/x$ being the exponential integral.
For the initial condition $\rho_{z}(0)=1$ the relaxation dynamics is then given by
\begin{multline}
P(t)= G^R(t,0)G^A(0,t)+\\
+ i \iint G^R(t,t_1)\Sigma^K(t_1,t_2)G^A(t_2,t)dt_{1}dt_{2},
\label{eq:PKel}
\end{multline}
with $G^{R/A}(t,t')=\mp i e^{\mp\int_{t'}^{t} d\tau \Gamma_0(\tau)/2}\theta(\pm[t-t'])$ 
and the self-energy $\Sigma^K(t_1,t_2)=\Gamma_0(t_1,t_2)/(\pi[t_2-t_1])$. 
We return to this equation when analyzing the behavior for small and asymptotically 
large $t$. In the former case the time dependent hybridization $\Gamma_0(t)=4|t_H(t)|^2/\omega_c$ 
matters, while in the latter 
one must also take into account $\Gamma_0(t,t')=4t_H(t)t_H^\ast(t')/\omega_c$. 

\subsubsection{Dynamics on short times}

In the RTRG method, for short times, such that $T_K t\ll 1$, the integral in 
Eq.~\eqref{eq:Prelax} is dominated by the large values $|E|\gg |\Gamma_{1}|$.
Therefore one may use the approximation $\Gamma_1(E)\approx T_{K}(-iE/T_{K})^{-g}$ 
of Eq.~(\ref{eq:solution_high_energies}) for the relaxation rate, which, since no information
on the infrared cutoff is left, is equivalent to the poor man scaling 
approximation \cite{Borda2007,Borda2008} here generalized to the case where the
Laplace variable serves as a cutoff parameter. 
Expanding the integrand of Eq.~(\ref{eq:Prelax}) in $\Gamma_1(E)/E$ and defining the variable
$z=Et$ gives the series
\begin{align}
\nonumber
P(t)&\approx \sum_{n=0}^\infty (-1)^n
(T_K t)^{n(1+g)}\int_{\mathcal{C}}\! e^{-iz}
(-iz)^{-1-n(1+g)} \frac{dz}{2\pi} \\
\label{eq:shorttimesRTRG}
&=\,\sum_{n=0}^\infty (-1)^n \frac{(T_{K}t)^{n(1+g)}}{\Gamma(1+n[1+g])},
\end{align} 
where $\Gamma(x)$ is the gamma function. With $g\equiv 1-2\alpha$ this result 
coincides precisely with the short time dynamics obtained using the noninteracting blip approximation (NIBA) 
for the SBM.\cite{Leggett87,Weiss12} However, one should keep in
mind that $\Gamma_1(E)\approx T_{K}(-iE/T_{K})^{-g}$ is only a good approximation for large $E$, i.e. only
the first two terms of the series Eq.~(\ref{eq:shorttimesRTRG}) can be trusted. Note that if one
neglects $g$ in the denominator $\Gamma(1+n(1+g))\approx\Gamma(1+n)=n!$, we obtain the
series of the exponential function and the result can be written as
\begin{align}
\label{eq:pms_short_times}
P(t)\,\approx\,e^{-(T_K t)^g\,T_K t} .
\end{align}
This coincides with the perturbative result $P(t)=e^{-\Gamma_t \,t}$, where 
$\Gamma_t=\Gamma(i/t)=T_K (T_K t)^g$ is the poor man scaling solution cut off 
at $-iE=1/t$. Solving the poor man scaling RG equation along the imaginary axis, i.e.
for $E=i\Lambda$, we see that $\Lambda$ serves as a flow parameter which has to be
cut off at $1/t$. Our finding is consistent with the generic result that the short-time behavior
probes the high-energy one in Laplace space and, therefore, can be obtained 
from poor man scaling equations cut off at the energy scale $1/t$, see Refs.~\onlinecite{Pletyukhov10,Kashuba13}.

\begin{figure}
\centering
\includegraphics[width=.9\columnwidth,clip]{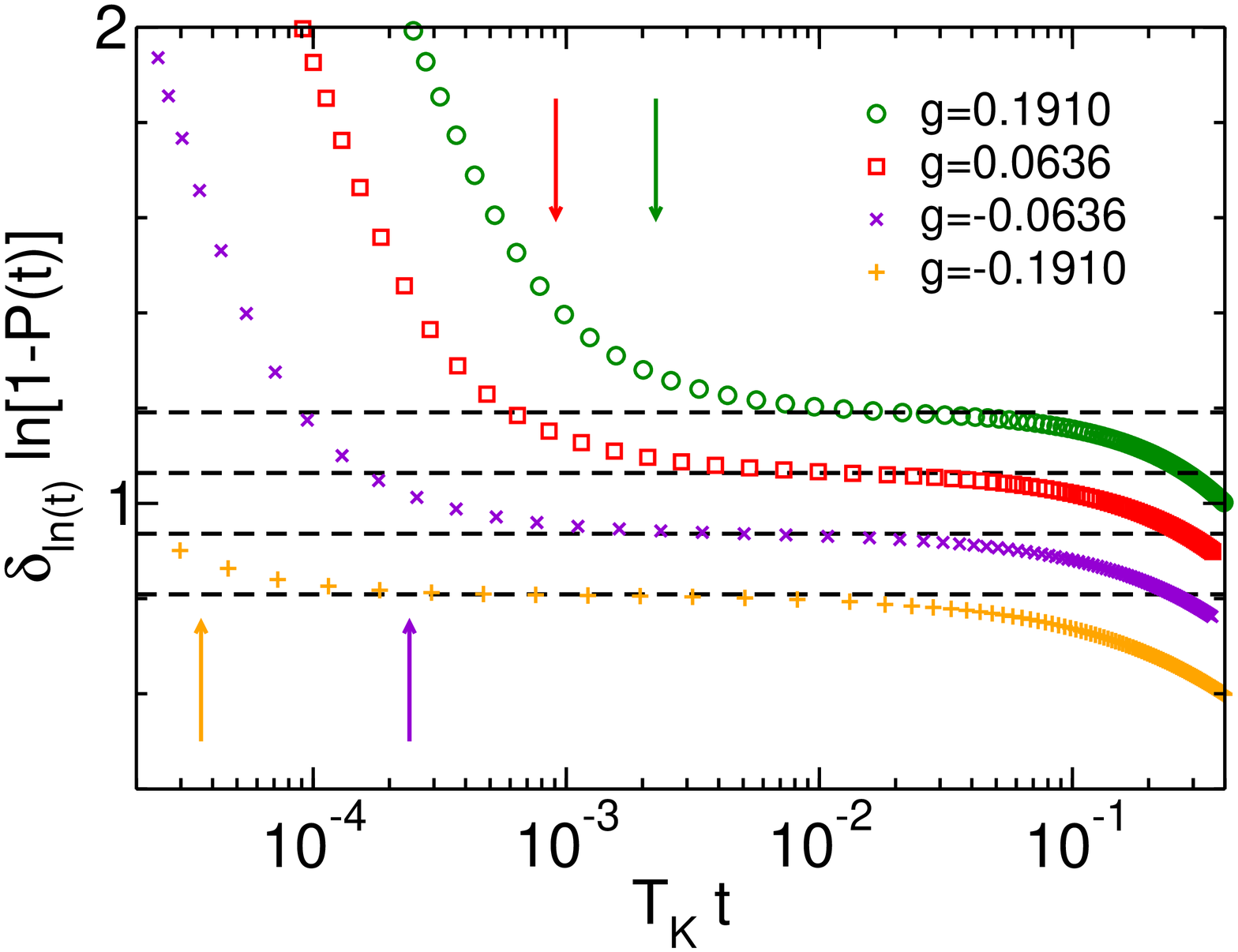}
\caption{(Color online) 
Power law scaling of $1-P(t)$ for $T_K t \ll 1$ in the relaxation protocol. 
The data are 
produced using the numerical solution of the full FRG flow equations.
Symbols represent the time dependence of the logarithmic derivative 
for different values of $g$. Horizontal dashed lines show the prediction 
$1+g$ of Eq.~\eqref{eq:shorttimesFRG} for the log-derivative (for the definition
of this, see the main text).
For $T_K t \lessapprox 10 T_K/\omega_c$ band effects start to matter and deviations from power-law scaling 
appear.  The arrows indicate the $g$-dependent $10 T_K/\omega_c$ 
(with $\omega_c/t_h=2\cdot 10^{2}$). 
}
\label{fig:logdershortt}
\end{figure}

Up to order $n=1$ the same result can be obtained from the FRG approach.
For times $\omega_c\gg 1/t\gg T_K$ the second term in Eq.~\eqref{eq:PKel} can be neglected 
and the argument of the exponential integral is small. We can thus replace 
${\rm E}_1(x)=-\gamma-\log(x)$, which leads to 
\begin{equation}
\Gamma_0(t)=T_K e^{\gamma g}\left(T_K t\right)^g
\end{equation}
for the time dependent renormalized hopping. Here 
$\gamma$ denotes the Euler constant. 
Using this the full retarded Green function at small times is given by
\begin{equation}
G^R(t,0) \approx -i \left(1-\frac{e^{\gamma g}}{2(g+1)}(T_K t)^{g+1}\right) ,
\end{equation}
leading to
\begin{equation}
P(t)=1-\frac{e^{\gamma g}}{g+1}(T_K t)^{g+1} +\mathcal{O}([T_K t]^2).
\label{eq:shorttimesFRG}
\end{equation}
Here, the power law with interaction dependent exponent is resummed correctly up 
to $\mathcal{O}(g)$ and the prefactor is determined within the same order.

The power-law scaling of $1-P(t)$ at $T_K t \ll 1$ can also be shown for 
the numerical data obtained by solving the full FRG flow equations (without any
additional approximations aside from the lowest-order truncation). To this
end, we numerically compute $\delta_{\ln(t)} \ln[1-P(t)] = 
d \ln[1-P(t)]/ d \ln t$ as centered differences, which becomes a 
constant if  $1-P(t)$ is given by a 
power law. In Fig.~\ref{fig:logdershortt}, we 
show $\delta_{\ln(t)} \ln[1-P(t)]$ for different $g$ as a function of $t$ on a log-linear 
scale. For $T_K t \approx 10^{-2}$ the data (symbols) become constant and nicely agree with
the exponent predicted by 
Eq.~\eqref{eq:shorttimesFRG} (dashed lines). The deviations for  
$t \lessapprox 10/\omega_c$ ($=5\cdot 10^{-2}/ t_h$ in the figure; the arrows indicate 
$10 T_K/\omega_c$) are an effect of the reservoirs band width; for such times
we leave the scaling limit. Further increasing 
$\omega_c$ in the numerical calculations would move this lower bound to smaller 
$t$.  

The comparison of the results obtained by our two RG approaches and the consistency 
with the results derived using established methods shows that for small times 
$T_K t \ll 1$ both methods provide controlled access to the relaxation dynamics.

\subsubsection{Intermediate to long times}

\label{sec:itolt}

The relaxation dynamics for times of the order of the typical inverse rates of the system and 
larger is of particular interest, as it indicates the degree of coherency.
For $t \to \infty$ the dot level becomes half-filled, that is the spin expectation 
value in the $z$ direction of the SBM vanishes. We say that the system is incoherent if 
it exhibits a monotonic decay for times $T_K t \gtrapprox 1$. Nonmonotonicity of $P(t)$, 
in particular zeros, is characteristic for coherent behavior. 

\begin{figure}
\centering
\includegraphics[scale=1]{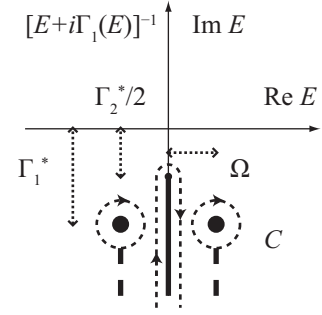}
\caption{The nonanalyticities of the propagator $[E+i\Gamma_{1}(E)]^{-1}$ in 
Eq.~\eqref{eq:Prelax} as a function of the complex variable $E$ for positive 
coupling $g>0$. The main branch cut (thick solid line) and two poles (circles) with attached 
second order in $g$ branch cuts (thick dashed lines) are shown.
The thin dashed line shows the integration contour $\mathcal C$ of 
Eq.~\eqref{eq:Prelax}.}
\label{fig:analytic}
\end{figure}

Within RTRG, the dynamics is determined via Eq.~(\ref{eq:Prelax}), given by
$P(t)=\frac{i}{2\pi}\int_{\mathcal{C}}dE e^{-iEt}R_1(E)$, and follows from 
the nonanalytical features of the resolvent $R_1(E)=1/[E+i\Gamma_1(E)]$ in the
lower half of the complex plane, see Fig.~\ref{fig:analytic}. We find three 
singularities parametrized by
\begin{equation}
\label{eq:z_i}
z_0\,=\,-\,i\,\frac{1}{2}\,\Gamma_2^* ,\quad
z_\pm\,=\,\pm\,\Omega\,-\,i\,\Gamma_1^* , 
\end{equation}
where the decay rates $\Gamma_i>0$ are of the order of $T_K$, whereas the 
oscillation frequency $0<\Omega\sim g T_K$.
The singularity $z_0$ is a branching point followed by a branch cut with
a discontinuous ${\mathcal O}(g)$ jump. We choose the position of the branch cut on 
the negative imaginary axis which turns out to be the most convenient choice 
to calculate the inverse Laplace transform for the time evolution (see below). 
$z_\pm$ denote the positions of two poles which are followed by two branch cuts
with an ${\mathcal O}(g^2)$ jump (which can be neglected in leading order). 
The pole and branch cut nonanalyticities lead to two terms in 
the time evolution which we denote by
\begin{equation}
\label{eq:P_splitting}
P(t)\,=\,P_{\text{pole}}(t)\,+\,P_{\text{bc}}(t) .
\end{equation}
The degree of coherence of $P(t)$ is given by the interplay of those  terms. 

An analytical understanding of the nonanalytical features of $R_1(E)$ can be obtained
by studying the RG equations (\ref{eq:rgflow}). We first see from the RG equation for $\Gamma_2(E)$
that if $E$ approaches a pole $z_\pm=-i\Gamma_1(z_\pm)$ of $R_1(E)$, 
$\Gamma_2(E)$ obtains a branch cut with jump of ${\mathcal O}(g)$ starting at $z_\pm$, which, when inserted
into the RG equation for $\Gamma_1(E)$ leads to a branch cut for $\Gamma_1(E)$ with jump of 
${\mathcal O}(g^2)$.
In addition, when $E$ approaches the pole $z_0=-i\Gamma_2(z_0)/2$ of $R_2(E)$, we see from the
RG equation for $\Gamma_1(E)$ that $\Gamma_1(E)$ has a branch cut starting at $z_0$ with jump 
of ${\mathcal O}(g)$. To calculate the positions of the singularities up to ${\mathcal O}(g)$, we solve
the RG equations by a systematic weak coupling expansion in $g\ll 1$. For $|E-z_i|\sim {\mathcal O}(T_K)$
and $g|\ln(|E-z_i|/T_K)|\ll 1$ we can expand the solution of Eq.~\eqref{eq:rgflow} in 
$g$ and can fix the integration constants by comparing with the solution
\eqref{eq:solution_high_energies} at high energies. Up to ${\mathcal O}(g)$ we obtain
\begin{align}
\label{eq:gam1_int_en}
\Gamma_1(E)/T_K\,&\approx\,1\,-\,g\,\ln\frac{-iE\,+\,\Gamma_2(E)/2}{T_K} ,\\
\label{eq:gam2_int_en}
\Gamma_2(E)/T_K\,&\approx\,1\,-\,g\,\ln\frac{-iE\,+\,\Gamma_1(E)}{T_K} .
\end{align}
Since $g|\ln{g}|\ll 1$ for $g\ll 1$, we can use these equations
for all $g^2\lesssim |E-z_i|/T_K \sim {\mathcal O}(1)$. This allows for a determination of
the positions of $z_i$ up to ${\mathcal O}(g)$.  We note that $\Gamma_{1/2}(E)$ has not been replaced
by $T_K$ in the argument of the logarithm since it is \emph{a priori} not clear whether
the argument becomes a negative real number of ${\mathcal O}(1)$, i.e., the ${\mathcal O}(g)$ correction 
of the imaginary part of $\Gamma_i(E)$ might be important. However, if $E$ is 
close to one of the pole positions $iz_\pm=\Gamma_1(z_\pm)$ or $iz_0=\frac{1}{2}\Gamma_2(z_0)$,
we can neglect this effect for the corresponding decay rate since we can expand around the 
pole position and use $\partial_E\Gamma_i\sim {\mathcal O}(g)$:
\begin{align}
\nonumber
\ln[-iE+\Gamma_2(E)/2]& \\
\nonumber
&\hspace{-1.5cm}
\approx\,\ln \left\{ -i(E-z_0)\left[1+i\partial_E\Gamma_2(z_0)/2\right] \right\}\\
\label{eq:gam2_log_exp}
&\hspace{-1.5cm}
\approx\ln[-i(E-z_0)] ,\\
\nonumber
\ln[-iE+\Gamma_1(E)]& \\
\nonumber
&\hspace{-1.5cm}
\approx \ln \left\{ -i(E-z_\pm)\left[1+i\partial_E\Gamma_1(z_\pm) \right] \right\} \\
\label{eq:gam1_log_exp}
&\hspace{-1.5cm}
\approx\ln[-i(E-z_\pm)] .
\end{align}
According to Eqs.~(\ref{eq:gam1_int_en}) and (\ref{eq:gam2_int_en}) this gives for $E=z_i+i\Lambda$, 
with $0<\Lambda\sim {\mathcal O}(g^2)$, the real value $1-g\ln(\Lambda/T_K)$ for $\Gamma_1(E)/T_K$ ($i=0$) 
or $\Gamma_2(E)/T_K$ ($i=\pm$).
Therefore, to determine the pole positions $z_0$ or $z_\pm$, we set $\Gamma_1(E)=T_K$ in 
Eq.~(\ref{eq:gam2_int_en}) or $\Gamma_2(E)=T_K$ in Eq.~(\ref{eq:gam1_int_en}), respectively, and obtain
with $-iz_0/T_K=-\frac{1}{2}+{\mathcal O}(g)$ and $-iz_\pm/T_K=-1+{\mathcal O}(g)$ for 
the ${\mathcal O}(g)$ correction:
\begin{align}
\nonumber 
2i z_0/ T_K &= \Gamma_2(z_0)/T_K 
\approx  1-g\ln(-iz_0/T_K+1) \\
\label{eq:z_0_determination}
&\approx  1-g\ln \left( -\frac{1}{2}+1 \right)  \approx 1+g\ln{2} \approx 2^g  ,\\
\nonumber 
i z_\pm/T_K &= \Gamma_1(z_\pm)/T_K 
\approx  1-g\ln \left( -iz_\pm/T_K+\frac{1}{2} \right) \\
\nonumber
&\approx  1-g\ln \left( -1\mp i\Omega/T_K+\frac{1}{2} \right) \\
\label{eq:z_+-_determination}
&\approx 1+g\ln{2}\pm i\pi g \approx 2^g \pm i\pi g  .
\end{align} 
Since $\Omega$ must be positive we see that the equation $E+i\Gamma_1(E)=0$ has only
a solution for positive coupling $g>0$. We note that $\Omega=0$ is excluded since
$E=-i\Gamma_1^*$ lies on the branch cut of $\Gamma_1(E)$. In conclusion, we obtain from 
Eqs.~(\ref{eq:z_0_determination}) and (\ref{eq:z_+-_determination})
the following result for the decay rates and the oscillation frequency up to ${\mathcal O}(g)$:
\begin{equation}
\label{eq:decay_rates_omega_O(g)}
\Gamma_{1/2}^*\,\approx\,2^g\,T_K ,\quad 
\Omega\,\approx\,\pi\,g\,T_K ,
\end{equation}
and we find that the poles at $z_\pm$ exist only for $g>0$.

If the Laplace variable $E$ lies exponentially close to one of the poles,
i.e. $g\ln(|E-z_i|/T_K)\sim {\mathcal O}(1)$, which probes the regime of exponentially large times
$g\ln(T_K t)\sim {\mathcal O}(1)$, we can replace $\Gamma_i(E)\rightarrow\Gamma_i(z_i)$ on the 
r.h.s. of the RG equations (\ref{eq:rgflow}). Fixing the integration constants
by comparison with the solutions \eqref{eq:gam1_int_en} and \eqref{eq:gam2_int_en} at intermediate
energies, we obtain
\begin{align}
\nonumber
&\underline{E\approx z_0}:\\
\label{eq:small_en_z0}
&\Gamma_1(E)\approx T_K\left[\frac{T_K}{-i(E-z_0)}\right]^g\,\,,\,\,\Gamma_2(E)\approx 2iz_0 ,\\
\nonumber
&\underline{E\approx z_\pm}:\\
\label{eq:small_en_z+-}
&\Gamma_1(E)\approx iz_\pm\,\,,\,\,\Gamma_2(E)\approx T_K \left[1-g\ln\frac{-i(E-z_\pm)}{T_K} \right] .
\end{align}
We note that the solutions \eqref{eq:solution_high_energies}, \eqref{eq:gam1_int_en},
\eqref{eq:small_en_z0}, and \eqref{eq:small_en_z+-} for $\Gamma_1(E)$ at high, intermediate, and 
exponentially small distances from $z_i$ can, in leading order in $g$, be interpolated by 
the compact expression
\begin{equation}
\label{eq:gam1_solution_all_en}
\Gamma_1(E)\,\approx\,T_K\,\left[\frac{T_K}{-i(E-z_0)}\right]^g .
\end{equation} 
One can check numerically that, for the special values $E=z_0+i\Lambda\pm 0^+$, with $\Lambda$ real, 
this formula holds even for larger values of $g$. Therefore, we will use this result below for the evaluation of
the branch cut integral to obtain the incoherent part $P_{\text{bc}}(t)$ of the time evolution. 
For $E$ exponentially close to the poles $z_\pm$ a numerical
analysis shows that Eq.~(\ref{eq:gam1_solution_all_en}) is correct for small values of $g$. For
larger couplings, it turns out that an improved fit is obtained by using
\begin{equation}
\label{eq:gam1_solution_z+-}
\underline{E\approx z_\pm}:\quad
\Gamma_1(E)\,\approx\,T_K\,\left[\frac{T_K}{-i(E-\Gamma_1(E)/2)}\right]^g .
\end{equation} 
This equation can be employed for an improved evaluation of the pole position and the residuum of
the resolvent $R_1(E)$. A straightforward calculation gives the result
\begin{align}
\label{eq:gamma_1*}
\Gamma_1^*/T_K\,&=\,2^\frac{g}{1+g}\, \left[1\,+\,\tan^2\left(\frac{\pi g}{1+g}\right)\right]^{-1/2}\,,\\
\label{eq:omega}
\Omega\,&=\,\Gamma_1^*\,\tan\left(\frac{\pi g}{1+g}\right)\,=\,
\Gamma_1^*\,\cot\left(\frac{\pi}{2}\,\frac{1-g}{1+g}\right)\,,\\
\label{eq:residuum}
&\hspace{-0.5cm}
\frac{1}{1+\partial_E\Gamma_1(z_\pm)}\,=\,\frac{1-g}{1+g}\,.
\end{align}
The ratio between the oscillation frequency $\Omega$ and the 
decay rate $\Gamma_1^*$, the so-called quality factor, was earlier 
computed using improved NIBA (Ref.~\cite{Egger97}) and a field theoretical 
approach.\cite{Lesage98} Our result is consistent with the one obtained 
within those approaches. 
Employing Eqs.~(\ref{eq:gamma_1*}) and (\ref{eq:residuum}) we
obtain directly from Eq.~(\ref{eq:Prelax})
\begin{equation}
\label{eq:prelax_coh}
P_{\text{pole}}(t)\,=\,2\,\frac{1-g}{1+g}\,\cos\left(\Omega t \right)\, 
e^{-\Gamma_1^\ast t}\,\Theta(g) .
\end{equation}

For $\Gamma_2(E)$ it is more difficult to find interpolation formulas valid for all values of
$E$ since this function behaves as a power law for high energies but like a logarithm for 
$E$ close to $z_\pm$. Inserting Eq.~(\ref{eq:gam1_solution_all_en}) into the RG equation 
(\ref{eq:rgflow}) for $\Gamma_2(E)$, approximating $\Gamma_1(E)\rightarrow z_\pm$ for 
$\text{sign}(\text{Re}[E])=\pm$, and taking another derivative, gives the differential 
equation for a special case of the hypergeometric function
\begin{equation}
\label{eq:hypo_1}
y(1-y)\frac{d^2\Gamma_2}{dy^2}+[1-(1+g)y]\frac{d\Gamma_2}{dy}=0\,,
\end{equation}
with $y=(z_\pm-E)/(z_\pm-z_0)$. In principle this equation can be solved providing an
interpolation formula for $\Gamma_2(E)$ valid for all $E$ and small $g$. However, 
since $\Gamma_2(E)$ is not needed for an evaluation of the time evolution of the diagonal 
matrix elements of the density matrix, we do not further discuss this issue. We note, that
$\Gamma_2(E)$ appearing in the resolvent $R_2(E)$ is important for the dynamics of
the off-diagonal matrix elements but those cannot be measured: It is impossible to
prepare an initial state which is off-diagonal in the charge states.

We next aim at the position of $z_0=-i\Gamma_2^*/2$ 
for larger values of $g$. Replacing $\Gamma_{1/2}(E)\rightarrow \Gamma_2^*$ on 
the r.h.s. of the RG equations (\ref{eq:rgflow}) provides a very good approximation as 
can be verified numerically. Based on this we show in Appendix~\ref{sec:app:rrc} how 
an improved solution can be found for $\Gamma_2(E)$ close to the imaginary axis 
for $E=z_0+i\Lambda$, with $\Lambda>0$. It leads to the improved formula 
\begin{equation}
\label{eq:gamma_2*}
\frac{\Gamma_2^*}{2}\,\approx\,\,T_K\,\left[\frac{\pi g}{2 \sin(\pi g)}\right]^\frac{1}{1+g}\,.
\end{equation}

We note that although the RG equations (\ref{eq:rgflow}) are only consistent up to ${\mathcal O}(g)$,
we have evaluated the analytical results Eqs.~(\ref{eq:gamma_1*}), (\ref{eq:omega}), and
(\ref{eq:gamma_2*}) for the position of the singularities taking higher-order corrections
in $g$ into account. This analysis is motivated by the fact that the ratio $\Omega/\Gamma_1^*$
agrees precisely with previous results from NIBA and field theoretical approaches which are
nonperturbative in $g$. Therefore, there is some hope that also the improved result 
Eq.~(\ref{eq:gamma_2*}) for $\Gamma_2^*$ is valid for larger values of $g$. 
To the best of our knowledge the rate
$\Gamma_2^*$ describing the energy broadening of the local state has so far
only been analyzed in leading order in $g$, see Ref.~\onlinecite{Egger97}.

Finally, we analyze $P_{\text{bc}}(t)$, given by 
the branch cut integral [we take $E=z_0-ix\pm 0^+$ with $0<x<\infty$ and
use $\Gamma_1(E)^*=\Gamma_1(-E^*)$]
\begin{align}
\nonumber
P_{\text{bc}}(t)\,&=\,\frac{1}{\pi}\,e^{-\frac{\Gamma_2^*}{2}t}\\
\label{eq:bc_integral_1}
&\hspace{-1cm}\times \text{Im}\int_0^\infty
\frac{e^{-xt}}{\Gamma_2^*/2 + x - \Gamma_1(z_0-ix-0^+)} dx ,
\end{align}
where $\Gamma_1(z_0-ix-0^+)\approx T_K (T_K/x)^{g}e^{-i\pi g}$ is taken 
from Eq.~(\ref{eq:gam1_solution_all_en}).
For intermediate to long times $T_K t\gtrapprox 1$ it is a very good approximation 
to replace the slowly varying function $x^g\rightarrow t^{-g}$ since the exponential
function $e^{-xt}$ in the integrand of Eq.~(\ref{eq:bc_integral_1}) restricts the integration
range to the regime $x\sim {\mathcal O}(1/t)$. This leads to the result
\begin{equation}
\label{eq:prelax_bc}
P_{\text{bc}}(t) \,\approx\, \frac{1}{\pi}\, 
\mbox{Im}\Bigl\{ e^{-\gamma_t t} \mbox{E}_1
\Bigl(\left[\tfrac{1}{2}\Gamma_{2}^\ast-\gamma_t\right] t\Bigr) 
\Bigr\} \,,
\end{equation}
with $\gamma_t = T_K (T_K t)^g e^{- i \pi g} $ and the exponential integral $E_1$. 

Combined with $P_{\text{pole}}(t)$ 
this result covers the limit $g\rightarrow 0$, where
$\gamma_t=T_K-i0^+\text{sign}(g)$ and 
$E_1(\Gamma_2^*/2-\gamma_t)= -\pi\,\text{sign}(g)$. This
gives together with Eq.~(\ref{eq:prelax_coh}) 
$P_{\text{bc}}(t)= -e^{-T_K t}\text{sign}(g)$ and 
$P_{\text{pole}}= 2e^{-T_K t}\theta(g)$, leading to the exact result
$P(t)=e^{-T_K t}$ and $T_K=\Gamma_0$ for $g=0$.

The contribution from the branch cut integral in the regime 
$1\lesssim T_K t \ll 1/g$ for small $|g| \ll 1$ can also be analyzed by observing that the
resolvent $R_1(z_0-ix+0^+)-R_1(z_0-ix-0^+)$ has a sharp Lorentzian peak at $x\sim T_K/2$
\begin{align}
\nonumber
R_1(z_0-ix+0^+)-R_1(z_0-ix-0^+)\,&\approx\,-2\pi\,\text{sign}(g) \\
\label{eq:lorentz_peak}
&\hspace{-6cm}
 \times \delta_{T_K(T_K t)^g\sin(\pi |g|)}\left(x+\frac{1}{2}\Gamma_2^*-T_K[T_K t]^g \cos[\pi g] \right) ,
\end{align}
where $\delta_\eta(x)=\frac{1}{\pi}\eta/(x^2+\eta^2)$. For $t\ll 1/g$, the peak width is
much narrower than the typical scale on which the function 
$e^{iEt}=e^{-\frac{1}{2}\Gamma_2^*t}e^{-xt}$ varies, i.e. we can replace the Lorentz peak 
by a $\delta$-function and get
\begin{align}
\nonumber
P_{\text{bc}}(t)\,&\approx\, -\,\text{sign}(g)\,e^{-T_K(T_K t)^g\cos(\pi g)t}\,\\
\label{eq:prelax_bc_small_t}
&\approx-\,\text{sign}(g)\,e^{-\Gamma_1^*t}\,.
\end{align}
Note that this leads to a contribution of ${\mathcal O}(1)$ and not of ${\mathcal O}(g)$.

In contrast, for long times $T_K t\gg 1/g$, we can use the asymptotic expansion 
$E_1(z)\approx e^{-z}/z$ and find from \eqref{eq:prelax_bc}
\begin{align}
\nonumber
P_{\text{bc}}(t) \,&\approx\, -\frac{\sin(\pi g)}{\pi} \\
\label{eq:prelax_bc_large_t}
&\hspace{-0.5cm}
\times \frac{1}{|\frac{1}{2}\Gamma_2^*/T_K-(T_K t)^g e^{i\pi g}|^2}\,
\frac{1}{(T_K t)^{1-g}}\,e^{-\frac{\Gamma_2^*}{2}t}\,\,.
\end{align}
Although this term is of ${\mathcal O}(g)$ it will dominate over $P_{\rm pole}(t)$
Eq.~(\ref{eq:prelax_coh}) of the time evolution as to leading order in $g$ its 
decay rate is only half of that of the coherent term.

For exponentially large times $(T_K t)^{|g|} \gg 1$, the branch cut contribution 
Eq.~(\ref{eq:prelax_bc_large_t}) can be approximated for $| g | \ll 1$ and $\Gamma_2^*\approx T_K$ by
\begin{equation}
P_{\text{bc}}(t)\,\approx\, - g [1+ 3 \Theta(-g)] \frac{e^{-T_K t/2}}{(T_K t)^{1+|g|}} .
\label{eq:ptstd}
\end{equation}
This result was earlier obtained using improved NIBA.\cite{Egger97} 
Note the power-law correction to the exponential decay already mentioned in the Introduction. 
We emphasize that 
this approximation is only justified for times so large that $P_{\text{pole}}(t)$ can safely 
be neglected compared to the dominating term $P_{\text{bc}}(t)$. Thus, the interesting 
competition of the branch-cut and pole contributions for $g>0$ on time scales of up to a few $T_K^{-1}$ 
can only be studied by keeping Eq.~\eqref{eq:prelax_bc}. In fact, it is this competition which for $g>0$ 
can lead to nonmonotonic, that is coherent, behavior of $P$ on time scales $T_{K}t\sim-\log g$ --- on 
asymptotic ones the dynamics is always monotonic due to the dominating branch cut contribution. 
On the coherent side of the coherent-to-incoherent transition but close to it the dynamics can thus 
only be classified as \emph{partially coherent}. This must be contrasted to the relaxation dynamics of the 
SBM at small spin-boson coupling $\alpha \ll 1$ and the one of the classical damped harmonic oscillator 
in its coherent regime (sufficiently weak damping) for which oscillatory behavior can be observed even on 
asymptotically large times.\cite{Leggett87,Weiss12,Anders06,Wang08,Orth10,Orth13} 
Analyzing the analytical expressions \eqref{eq:prelax_coh} and \eqref{eq:prelax_bc} for $P(t)$, one 
finds that up to a certain coupling $g_1>0$ only a single local minimum associated with a single transition 
through zero appears. For $g_1< g < g_2$, with a certain $g_2$, a second zero is found and so on. As our approach 
is limited to $|g| \ll 1$, we can only speculate about how this behavior crosses over to the coherent 
behavior with infinitely many zeros obtained at small $\alpha$. From the results \eqref{eq:gamma_1*}
and \eqref{eq:gamma_2*}, we suspect that beyond a characteristic coupling of order $g\sim 0.4$
[$\alpha=(1-g)/2\sim 0.3$] the finite frequency poles lie closer to the real axis than the branching 
point, and oscillatory dynamics dominates at large times.    

The discussed behavior is confirmed by the full numerical solution of the (approximate) FRG and 
RTRG flow equations. 
\begin{figure}
\centering
\includegraphics[width=.9\columnwidth,clip]{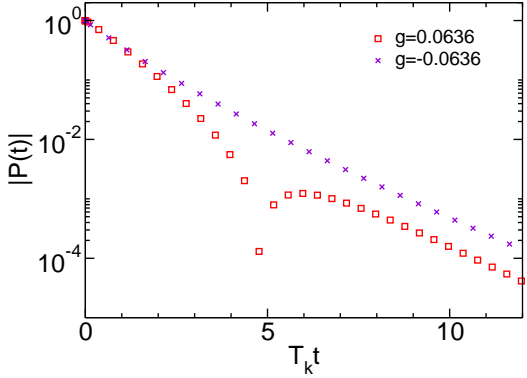}
\caption{(Color online) 
$|P(t)|$ for all time scales.
For $g>0$ the competition of the two Eqs.~\eqref{eq:prelax_coh} and \eqref{eq:prelax_bc} 
manifest in the sign change of $P(t)$, appearing as a dip on 
the logarithmic scale.
The partially coherent ($g>0$, red squares) and incoherent ($g<0$, purple crosses) 
dynamics are compared.}
\label{fig:solotime}
\end{figure}
In Fig.~\ref{fig:solotime}, we show $|P(t)|$ obtained from FRG for small 
$|g|$ on a linear-log scale. For $g>0$ the zero appears as a dip.  
For negative $g$ there are no poles, so $P_{\text{coh}}=0$. Analyzing Eq.~\eqref{eq:prelax_bc} 
shows that it contains two terms of the same sign both showing exponential decay with the different 
rates $\Gamma_{1}^\ast$ and $\Gamma_{2}^\ast/2$.  In this regime the dynamics is clearly 
incoherent; see the crosses in Fig.~\ref{fig:solotime}.
In Fig.~1 of Ref.~\onlinecite{Kennes13b} we present a detailed 
comparison of $P(t)$ obtained from the approximate analytical solution of the RTRG flow equations 
given in Eqs.~\eqref{eq:prelax_coh} and \eqref{eq:prelax_bc}, as well as the 
numerical solution of those and the FRG flow
equations. This figure also contains a data set from the regime $g_1< g < g_2$ showing two zeros.  

For $g>0$ the asymptotic long time behavior can also be accessed analytically using FRG. For those 
the second term of Eq.~\eqref{eq:PKel} is the dominant one, since it decays slower than the first 
one (see below).  As ${\rm{Im}}\left[P(t) \right]= 0$ the second term of Eq.~\eqref{eq:PKel} only has a 
nonzero contribution for 
${\rm{Im}}\left[\Sigma^K(t_1,t_2) \right]={\rm{Im}}\left\{4t_H(t)t_H^\ast(t')
/[\omega_c\pi(t_2-t_1)]\right\}\neq 0$. Therefore we concentrate on this imaginary part, where $t_H(t)$ is given by 
Eq. \eqref{eq:th_ren}, which leads to
\begin{align}
&{\rm{Im}}\left[\Sigma^K(t_1,t_2)\right]=T_K   {\rm{Im}}\Bigl\{\exp\Bigl[-\frac{i \pi g}{4}\Bigl( e^{-\Gamma_0 t_1} -e^{-\Gamma_0 t_2}\Bigr) \Bigr]\Bigr\} \notag\\
&\times \frac{1}{\pi(t_2-t_1)}+\mathcal{O}(g^2)\notag\\
=&T_K \, {\rm{Im}}\Bigl\{\exp\Bigl[-\frac{i \pi g}{4} e^{-\Gamma_0 T/2}\Bigl( e^{\Gamma_0 \Delta t/2} -e^{-\Gamma_0 \Delta t/2}\Bigr) \Bigr]\Bigr\} \notag\\
&\times \frac{1}{\pi \Delta t}+\mathcal{O}(g^2),
\label{tudel}
\end{align}
with $T=t_1+t_2$ and $\Delta t=t_2-t_1$.
The integrals in the second 
term of Eq.~\eqref{eq:PKel} are dominated by times $T_{K}(t-t_{1/2})\lesssim 1$ such that for 
$(T_K t)^{|g|} \gg 1$ one can also use $(T_K t_{1/2})^{|g|} \gg 1$ 
inside the integral.
For $t\to\infty$ we can thus replace $e^{-\Gamma_0 T/2}$ in Eq.~\eqref{tudel} by its time 
averaged value over $[t,2t]$ and using 
$\Gamma_0 \Delta t \propto \Gamma_0/T_K\ll 1$ find
\begin{align}
{\rm{Im}}\left[\Sigma^K(\Delta t; t)\right] = &\frac{T_K}{\pi \Delta t}{\rm{Im}}\left(\exp\left[-i\frac{\pi g}{2}\frac{\Delta t}{t}
e^{-\Gamma_0 t/2}\right]\right)\label{eq:SigmaKfinal}.
\end{align}
This self-energy is formally equivalent to one which arises due to a noninteracting reservoir held at 
chemical potential $\mu_t=-\frac{\pi g}{ 2t}e^{-\Gamma_0 t/2}$, where $t$ can be viewed as a 
parameter. Using this and $\Gamma_0/T_K\ll 1$ (which holds for $g>0$)  Eq.~\eqref{eq:PKel} for 
large times simplifies to   Eq.~\eqref{eq:ptstd}.
\emph{A posteriori} we can justify that we only considered the second term of Eq.~\eqref{eq:PKel} as 
it decays exponentially with rate $T_K/2$ while the first term leads to an exponential decay with rate 
$T_K $.

The consistency of all our analytical and numerical results for the relaxation dynamics 
for intermediate to large times and the agreement to established results in the cases in 
which a comparison is meaningful confirms that our two RG approaches provide controlled
access to the dynamics of the SBM close to the coherent-to-incoherent transition. 

Finally, we note that for small couplings $g\ll 1$, it is tempting to interpolate the results
\eqref{eq:prelax_bc_small_t} for $1\lesssim T_K t \ll 1/g$ and \eqref{eq:prelax_bc_large_t} for
$1/g \ll T_K t$ by taking the sum of both terms. Adding the pole contribution \eqref{eq:prelax_coh}
provides the following result for $P(t)$  
\begin{align}
\nonumber
P(t)\,&\approx\,e^{-\Gamma_1^* t}\,\left[2\cos(\Omega t)\theta(g)-\text{sign}(g)\right]\\
\label{eq:prelax_intermediate_t}
&-g\,\frac{1}{(\frac{1}{2}\Gamma_2^*/T_K-(T_K t)^g)^2}\,\frac{1}{(T_K t)^{1-g}}\,e^{-\frac{1}{2}\Gamma_2^* t}\,.
\end{align}
This formula can be used for $g\ll 1$ and $T_K t\gtrsim 1$ and covers the correct limiting cases 
of $T_K t\ll 1/g$ and $T_K t\gg 1/g$ since the second term is of ${\mathcal O}(g)$ but decays with a smaller decay rate.
Therefore, the first term will dominate for $T_K t\ll 1/g$ whereas the second one dominates for
$T_K t\gg 1/g$.
However, we note that this equation covers the intermediate time regime $T_K t\sim 1/g$ only
on a qualitative level. It will be very helpful for a qualitative
discussion of the quench protocol considered in the next section. 

\subsection{Quench protocols}
\label{sec:quench_protocols}

The dynamics resulting out of different types of parameter quenches in the IRLM was studied 
earlier,\cite{Anders06,Kennes12b,Eidelstein12,Guettge13} but no thorough 
analysis of the role of non-Markovian memory was presented so far. 

\subsubsection{The RTRG approach}
\label{sec:quench_rtrg_irlm}

We begin with adapting the general approach to quenches using the RTRG method described in Sect.~\ref{qRTRG}
to the IRLM. Inserting Eq.~\eqref{eq:sigma12} in Eq.~\eqref{eq:time_evolution_quench} and using the form
$R_{f/i}(E)=\tilde{R}_\Delta^{f/i}(E)Z_{f/i}^\prime(E)$ of the resolvent, we obtain for $t_0=0$
\begin{widetext}
\begin{align}
\nonumber
\rho_f(t)\,&=\,\frac{i}{2\pi} \int e^{-iE(t-t_q)} \,\tilde{R}^f_\Delta(E) \,Z_f^\prime(E) \,\rho(t_q) dE
\,-i\,\int_0^\infty  \iint e^{-iE(t-t_q)}e^{-iE't_q} \\
\label{eq:rho_quench_zw}
&\times \sum_{\eta=\pm}\,\tilde{R}_\Delta^f(E)\,\tilde{G}^f_\eta(E)
\,\tilde{R}_\Delta^f(E+i\Lambda)\,Z_f^\prime(E+i\Lambda)
\,\tilde{R}_\Delta^i(E'+i\Lambda)\,\tilde{G}^i_{-\eta}(E')\,\tilde{R}_\Delta^i(E')\,
Z_i^\prime(E')\rho(0)  \frac{dE dE'}{(2\pi)^2} d\Lambda,
\end{align}
\end{widetext}
where $\tilde{G}^{f/i}_\eta(E)=Z'_{f/i}G^{f/i}_\eta(E)$. According to Eq.~\eqref{eq:sigma_z},
the first term on the r.h.s. gives the following contribution to $\langle\sigma_z\rangle(t)$
\begin{equation}
\nonumber
P_f(t-t_q)\langle\sigma_z\rangle(t_q)=P_f(t-t_q)P_i(t_q)\langle\sigma_z\rangle(0) ,
\end{equation}
where $P_{i/f}(t)$ is the dot occupancy (spin expectation value) computed without a quench as given in 
Eq.~\eqref{eq:Prelax} using the parameters of $H_{\text{tot}}^i$ and $H_{\text{tot}}^f$, respectively. 
For the function $P(t)$, this gives the contribution $P_f(t-t_q)P_i(t_q)$.

The second term on the r.h.s. of Eq.~\eqref{eq:rho_quench_zw} can be evaluated by inserting the
matrix structure of $\tilde{R}_\Delta^{f/i}$, $Z_{f/i}^\prime$ and $\tilde{G}_\eta^{f/i}$, as provided in 
Appendix~\ref{sec:app:rtrg_irlm}. Since we are interested in the diagonal elements
of the density matrix, a straightforward analysis shows that only the following parts of the
various resolvents contribute:
\begin{align}
\nonumber
\tilde{R}_\Delta^{f/i}(E)\,&\rightarrow\,
\frac{1}{E+i\Gamma_1^{f/i}(E)}\,\left(\begin{array}{c|c}\tau_- & 0 \\ \hline 0 & 0 \\ \end{array}\right)\,,\\
\nonumber
\tilde{R}_\Delta^{f/i}(E+i\Lambda)\,&\rightarrow\,
\frac{1}{E+i\Lambda+i\Gamma_2^{f/i}(E+i\Lambda)/2}\,
\left(\begin{array}{c|c}0 & 0 \\ \hline 0 & \mathbbm{1} \\ \end{array}\right)\,.
\end{align}
Neglecting $\Lambda$ in the slowly varying function
$Z'(E+i\Lambda)$ and inserting the matrix structure of $Z'$ and the vertices $\tilde{G}^{f/i}_\eta$, we
find after straightforward algebra for the part determining the diagonal matrix elements of the
density matrix that in Eq.~\eqref{eq:rho_quench_zw} we can employ the replacement
\begin{widetext}
\begin{align}
\nonumber
&\sum_{\eta=\pm}\,\tilde{R}_\Delta^f(E)\,\tilde{G}^f_\eta(E)
\,\tilde{R}_\Delta^f(E+i\Lambda)\,Z_f^\prime(E+i\Lambda)
\,\tilde{R}_\Delta^i(E'+i\Lambda)\,\tilde{G}^i_{-\eta}(E')\,\tilde{R}_\Delta^i(E')\,
Z_i^\prime(E')\,\rho(0) \\
\nonumber
&\rightarrow\,-i\,2U_i\,\sqrt{Z_f(E)Z_i(E')\Gamma_1^f(E)\Gamma_1^i(E')}\,
R_1^f(E)\,R_2^f(E+i\Lambda)\,R_2^i(E'+i\Lambda)\,R_1^i(E')
\,\frac{1}{2}\,\begin{pmatrix}-1\\ 1\\\end{pmatrix}\,\langle\sigma_z\rangle(0)\,,
\end{align}
\end{widetext}
where $R_{n}^{f/i}(E)=1/[E+i\Gamma_n^{f/i}(E)/n]$.
As a result the contribution of Eq.~\eqref{eq:rho_quench_zw} to the function $P(t)$ can
be written as
\begin{subequations}
\begin{align}
P(t) &= P^{f}(t-t_q)P^{i}(t_q) \nonumber\\&- 
g_i \int_{0}^{\infty} F_{\Lambda}^{f}(t-t_q)F_{\Lambda}^{i}(t_q) d\Lambda,
\label{eq:PquenchI}
\end{align}
where $g_i\approx 2U_i$ in leading order, and the functions $F_{\Lambda}^{i/f}(t)$ are defined by
\begin{align}
F^{i/f}_{\Lambda}(t) &= \int e^{-iEt}  R_{2}^{i/f}(E+i\Lambda) \nonumber
\\
&
\times
R_{1}^{i/f}(E)\,\sqrt{Z_{i/f}(E)\,\Gamma_{1}^{i/f}(E)} \frac{dE}{2\pi}
\label{eq:PquenchF}
\end{align}
and describe the effective emission of a single fermionic excitation by the 
dot before the quench ($F^{i}$) and its absorption after the quench ($F^{f}$).
\label{eq:Pquench}%
\end{subequations}
The subsequent analytical evaluation of this formula in leading order in $g$ is shown in 
Appendix~\ref{sec:app:kern}. We note that this analysis is only qualitative, but in essence 
backed-up 
by the numerical solution of the full RG equations, since we need results in the 
intermediate-time regime either for $T_K^f(t-t_q)\gtrapprox 1/g_f$
or for $T_K^i t_q\gtrapprox 1/g_i$, where Eq.~\eqref{eq:prelax_intermediate_t} for
the relaxation protocol is only qualitative. However, these
interpolation formulas are sufficient for a discussion of the competition between 
the two terms of $P(t)$ in Eq.~\eqref{eq:PquenchI} describing the effects without and 
with memory from the dynamics before the quench.

\subsubsection{The FRG approach}

In a two-lead geometry, quenches of the single-particle parameters (but not the 
two-particle interaction $U$) of the IRLM were earlier studied in 
Ref.~\onlinecite{Kennes12b} using FRG. The focus was on systems driven by an applied 
bias voltage. Within the FRG approach also the interaction can
carry an arbitrary time dependence, such that the approach developed in this paper
can directly be applied to the problem of present interest.    

\subsubsection{Interaction quenches}

To illustrate the physics of a quench, which is substantially different from the 
one of the simple relaxation process, we initially study the instantaneous transition between the 
(partially) coherent ($g>0$) and the incoherent ($g<0$) regimes, keeping the absolute value 
$|g|$ fixed. The characteristic energy scales $T_K^{i/f}$ before and after the 
quench differ by orders of magnitude. As a consequence, the new small parameter 
$A^{2}$ appears which is given by the ratio of the smaller Kondo temperature at negative coupling 
and the larger one at $g>0$. Substituting the expressions for 
$T_K^{i/f}\approx\Gamma_0 (\omega_c/\Gamma_0)^{g_{i/f}}$ from Eq.~\eqref{eq:T_K} for $g\ll 1$ we 
obtain $A \approx(\Gamma_{0}/\omega_c)^{|g|}$. 

\paragraph{Coherent to incoherent quench.}

\begin{figure}
\centering
\includegraphics[width=.9\columnwidth,clip]{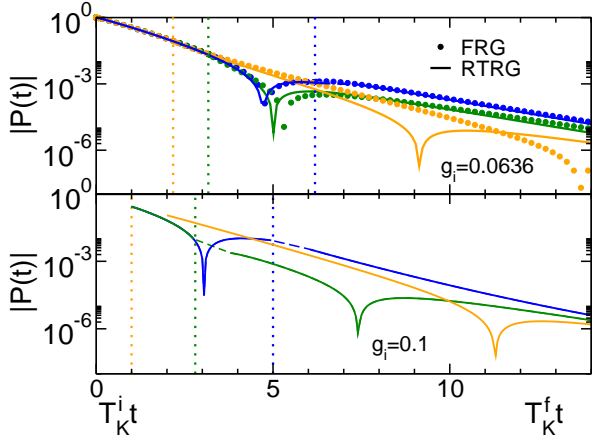}
\caption{(Color online) Time evolution of the spin expectation value $P(t)$ 
in the quench protocol. The coupling is quenched from $g_i > 0$ to $g_f = -g_i <0$. 
The time of the quench $t_q$ is varied as indicated by the vertical dotted lines. If $t_q$ 
is smaller than the time $t^*$ of the first zero of $P(t)$ in the absence of the 
quench the non-Markovian memory transfers the nonmonotonic behavior to the time 
regime after the quench in which the time evolution is performed with a negative 
coupling constant.
Note that the time axis is chosen to be $T_K^i t$ before the quench and   
$T_K^f t$ after it. Since $T_K^i$ and $T_K^f$ can differ by orders of magnitude
scaling the time axis in this way is physically meaningful.   
Upper panel: $P(t)$ obtained
from the numerical solution of the FRG (symbols) and RTRG (lines) flow equations. 
Lower panel: the analytical RTRG result Eq.~\eqref{eq:QGci:sub}. Curves are shown 
only for times $T_{K}^{f}(t-t_{q})\gtrsim1$  for which this result is applicable.}
\label{fig:QuenchGci}
\end{figure}

We first consider a quench from $g_i=g>0$ to $g_f=-g<0$.
The dynamics of the system in the (partially) coherent regime implies at least one local minimum and one 
zero of $P(t)$ in the absence of a quench. The position of the first zero of the relaxation protocol will 
be denoted as $t^*$ in the following. In the absence of the memory term from the quench, 
obtained within RTRG by leaving out the second term 
in Eq.~\eqref{eq:PquenchI}, $P(t)$ after the quench is monotonically decaying to zero. 
In the systems memory from the dynamics before the quench, that is the second term of  
Eq.~\eqref{eq:PquenchI}, however, 
the tendency to change the sign of $P(t)$, is kept even at $t>t_q$. Thus, if the quench is 
performed at time $t_{q} < t^*$, nonmonotonic behavior appears for $t > t_q$ for which the time evolution 
is already performed with a negative coupling.  
In Appendix~\ref{sec:app:kern} we show that Eq.~\eqref{eq:Pquench} can be written as
\begin{eqnarray}
P(t) &\approx&
e^{-\Gamma_{1}^{*f}(t-t_{q})-\Gamma_{1}^{*i}t_{q}}\bigl[2\cos(\Omega_{i} t_{q})-1\bigr] -
\nonumber\\
&-& g A S^{-}_{T_{K}^{f}t}S^{+}_{T_{K}^{i}t} 
\frac{e^{-\Gamma_{2}^{*f}(t-t_{q})/2-\Gamma_{2}^{*i}t_{q}/2}}{T_{K}^{f}t},
\label{eq:QGci:sub}
\end{eqnarray}
where \mbox{$A=\sqrt{T_{K}^{f}/T_{K}^{i}}$} and the  function $S^{\pm}$ is defined as
\begin{equation}
S^{\pm}_{x}=\frac{(x/2)^{\pm g/2}}{(x/2)^{\pm g}-1/2}.
\label{eq:Sdef}
\end{equation}
Note that we have included for $P^f(t-t_q)P^i(t_q)$ only the first term of
Eq.~\eqref{eq:prelax_intermediate_t}. The contributions of ${\mathcal O}(g)$ resulting from the 
second term of Eq.~\eqref{eq:prelax_intermediate_t} lead to a stronger decay compared
to the second term of \eqref{eq:QGci:sub}. The first term of Eq.~\eqref{eq:QGci:sub}, 
corresponding to the coherent contribution, and the second one, describing the 
system memory, have different signs for $t_{q}<t^\ast$, and therefore compete.
If $2\cos(\Omega_{i} t_{q})-1\gg g$, then, right after the quench the first term dominates, 
but, due to the twice larger decay rate, becomes recessive later. 
This leads to a local minimum and  
a sign change after the quench. With $t_{q}$ approaching $t^\ast$, $2\cos(\Omega_{i} t_{q})-1$ 
decreases and the time scale at which the second term dominates approaches $t_q$ from above.
For times $t_{q}>t^\ast$, the first one is already negative, so both terms 
just add up and no local minimum or zero of
$P(t)$ appears. In the lower panel of Fig.~\ref{fig:QuenchGci}, the absolute value of Eq.~\eqref{eq:QGci:sub} 
is shown on a linear-log scale for fixed $g_i=0.1$, different $t_q$, and the times of validity of this 
expression (see above). 

The analytical prediction of Eq.~\eqref{eq:QGci:sub} is confirmed by the numerical solutions of the FRG  
and RTRG flow equations. The corresponding $|P(t)|$ are shown in the upper panel 
of Fig.~\ref{fig:QuenchGci} (symbols: FRG; lines: RTRG). 
If $t_q$ (indicated by the vertical dotted lines) is larger than $t^*$ (given by the dip position
of the blue curve with $t_q$ to the right of the dip), $P(t)$ after the quench does 
not show a local minimum and/or
zero (blue curve). In the opposite case  (green and yellow) nonmonotonic behavior is 
transferred deep into the regime in which the time evolution 
is performed with a negative $g$. For the case in which $t_q$ is significantly to the left of $t^*$ 
(yellow curve) the FRG and the RTRG data show sizable differences at 
large $t$. In this case the zero of $P(t)$ is transferred to very large times, that is deeply 
into the regime in which the time evolution is performed with a negative coupling constant. At 
those times the overall exponential decay already suppressed $|P(t)|$ to a very small value (of the
order $10^{-6}$ in the figure). Therefore any small difference in the $P(t)$ obtained by the two 
approximate methods leads to a drastically different appearance of the curves (e.g.~a sizable shift 
of the zero) on the  linear-log scale of the plot.

\paragraph{Incoherent-to-coherent quench.}

\begin{figure}
\centering
\includegraphics[width=.9\columnwidth,clip]{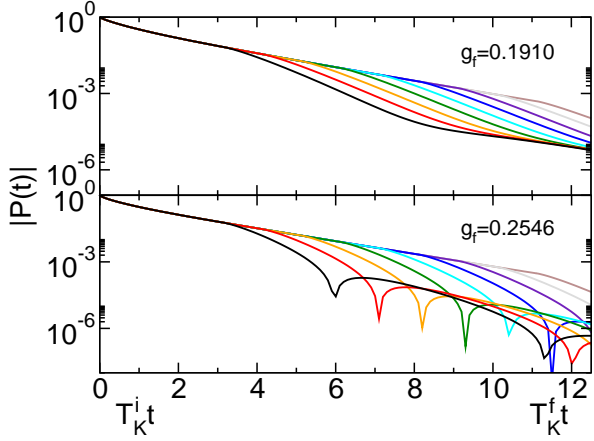}
\caption{(Color online) Time evolution of the spin expectation value $P(t)$  
in the quench protocol obtained from the numerical solution of the FRG flow equations. 
Here the coupling is quenched from $g_i < 0$ to $g_f = -g_i >0$. 
Note that the time axis is chosen to be $T_K^i t$ before the quench and   
$T_K^f t$ after it. Different lines correspond to different quench times $t_q$ increasing 
from left to right. Upper panel: $g_f < g_c$, that is the time evolution even after the quench
is monotonic even though it is performed with a positive coupling constant. 
The non-Markovian memory heavily affects the dynamics after the quench.
The qualitative behavior  is independent of $t_q$. Lower panel: $g_f > g_c$; nonmonotonic behavior
is found after the quench. This is again independent of $t_q$.}
\label{fig:QuenchGic}
\end{figure}

We next study the opposite interaction quench with $g_i=-g<0$ to $g_f=g>0$ and 
show that monotonic behavior may prevail even after the quench. Again 
the non-Markovian memory is responsible for this surprising behavior.
To gain analytical insights we consider Eq.~\eqref{eq:Pquench} 
in the limit $T_{K}^{i}t_{q}\gg T_{K}^{f}(t-t_{q})\gtrsim 1$. 
As described in detail in Appendix~\ref{sec:app:kern}, we obtain 
the approximate expression 
\begin{eqnarray}
&&\displaystyle\frac{P(t)}{P(t_{q})} =
\left[ 1\!-\!\frac{2 A}{S^{-}_{T_{K}^{i}t_{q}}} \right] e^{-\Gamma_{1}^{*f}(t-t_{q})}
\nonumber\\&&\times \!
\left[ 2\cos\left\{\Omega_{f}(t\!-\!t_{q})\right\} \!-\! 1 \right] \!  + \! 
A \frac{S^{+}_{T_{K}^{f} t_{q}}}{S^{-}_{T_{K}^{i}t_{q}}}  e^{-\Gamma_{2}^{*f}(t-t_{q})/2} ,
\label{eq:QGic:sub}
\end{eqnarray}
with \mbox{$A=\sqrt{T_{K}^{i}/T_{K}^{f}}$}. We left out the second term
of Eq.~\eqref{eq:prelax_intermediate_t} for the calculation of $P^f(t-t_q)$ in the first
term of Eq.~\eqref{eq:PquenchI} since it leads to a subleading contribution 
${\mathcal O}(g_f)$ to Eq.~\eqref{eq:QGic:sub}. We are not interested in exponentially 
large times and can thus set $S^{\pm} = 2$ in Eq.~\eqref{eq:QGic:sub}. It simplifies to
\begin{eqnarray}
\displaystyle\frac{P(t)}{P(t_{q})}&\approx&
(1-A) e^{-\Gamma_{1}^{*f}(t-t_q)}\left[2\cos\left\{\Omega_{f}(t-t_{q})\right\} - 1 \right]
\nonumber\\&&+
A\, e^{-\Gamma_{2}^{*f}(t-t_q)/2}.
\label{eq:QGic}
\end{eqnarray}
This expression shows that because of $A\approx(\Gamma_{0}/\omega_c)^{|g|}$,  
the oscillatory term of the dynamics after the quench is suppressed with decreasing $g$. A 
critical $g_{c}$ is found such  that for $g<g_{c}$ the relaxation is monotonic after the quench,
even though the time evolution is performed with a positive coupling constant.  
This result is again confirmed by FRG calculations revealing the independence of the qualitative
behavior of $P(t)$ from the time of the quench $t_q$ (see Fig.~\ref{fig:QuenchGic}). We here 
refrain from showing results obtained from the numerical solution of the RTRG equations and just
mention that those are consistent with the behavior described above.

The transfer of monotonic dynamics across the quench can be explained as follows.
In the case of the relaxation protocol with $g>0$ there are two competing terms with different 
relaxation rates and prefactors, which result from the pole 
$\sim e^{-\Gamma_{1}^{*f}(t-t_{q})}[2\cos\{ \Omega_{f}(t-t_{q})\}-1]$ and branch cut 
$\sim g e^{-\Gamma_{2}^{*f}(t-t_{q})/2}$ contributions.
The latter follows from the non-Markovian terms in the von Neumann equation, and 
therefore represents the memory of the system.
For the sake of simplicity of the qualitative explanation we here use the simplified 
expressions omitting the power-law dependencies.
Due to the small prefactor $g$, the oscillating term always dominates at times 
$T_{K}^{f}(t-t_{q})\sim1$, but due to the twice larger decay rate is subdominant at 
larger times. After the quench, the memory term remains the same, 
i.e., $\sim g e^{-\Gamma_{2}^{*f}(t-t_{q})/2-\Gamma_{2}^{*i}t_{q}/2}$,  where the exponent contains the 
contributions collected starting at $t=0$. The contribution without memory from the
dynamics before the quench, which is equal to 
$e^{-\Gamma_{1}^{*f}(t-t_{q})}[2\cos\{\Omega_{f}(t-t_{q})\}-1]$, 
where $t-t_{q}$ is the time passed after the quench, 
needs to be multiplied by the value of the density matrix at $t_{q}$, which is 
$g e^{-\Gamma_{2}^{*i}t_{q}/2}$. Thus, the competing terms after the quench are 
$\sim g e^{-\Gamma_{1}^{*f}(t-t_{q})-\Gamma_{2}^{*i}t_{q}/2} [2\cos\{\Omega_{f}(t-t_{q})\}-1]$ 
and $\sim g e^{-\Gamma_{2}^{*f}(t-t_{q})/2-\Gamma_{2}^{*i}t_{q}/2}$, and, depending on 
more subtle prefactors [see Eq.~\eqref{eq:QGic}], the nonoscillatory term may 
dominate even right after the quench.

\subsubsection{Quench of the level position}

\begin{figure}
\centering
\includegraphics[width=.9\columnwidth,clip]{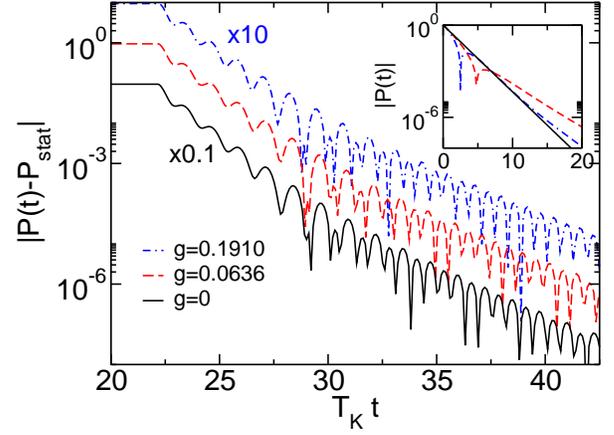}
\caption{(Color online) Time evolution of the spin expectation value $P(t)$ after a quench of the Zeeman splitting 
from zero to $\epsilon/T_K=5$ obtained within FRG. The stationary value $P_{\rm stat} \neq 0$ was 
subtracted and the data are presented on a linear-log scale.  The behavior for systems with different 
couplings [$g=0.0636$ (dashed red line) and $g=0.1910$, 
(dotted-dashed blue line)] does not significantly differ from the noninteracting case (black solid line).
The inset shows $|P(t)|$ before the quench.} 
\label{fig:QuenchEps1}
\end{figure}

Up to now we exclusively considered the case of vanishing level energy $\epsilon=0$  or, for the SBM, vanishing
Zeeman splitting in both the relaxation as well as the quench protocols. We next study the response of the 
system to abruptly changing $\epsilon$ for $g \geq 0$. 
For this we only use the flexible FRG approach in which finite $\epsilon$ can straightforwardly 
be included.  
We start out with the case of a quench from $\epsilon=0$  
to a value much larger than $T_K$. 
The instantaneous increase of the level position from zero, that is resonance, to a large value results 
in a behavior of $P(t)$ similar to the one observed in the relaxation protocol with a time independent 
large $\epsilon$; for details see Refs.~\onlinecite{Andergassen11} and \onlinecite{Kennes12}. For a 
finite Zeeman field the spin expectation value does no longer vanish at large times (the level 
occupancy does not approach $1/2$). In Fig.~\ref{fig:QuenchEps1} showing $P(t)$ for different 
$g$ we thus subtracted the stationary value $P_{\rm stat}$. Qualitatively the dynamics can be described 
by considering two main contributions. The nonoscillatory term, $e^{-\Gamma_{1}^{*}(t-t_{q})}$ stemming 
from a pole on the imaginary axis and the term $(T_{K}/\epsilon)e^{-\Gamma_{1}^{*}(t-t_{q})t/2}
\sin[\epsilon (t-t_{q})]/[\epsilon (t-t_{q})]$ originating from a pair of branch cuts positioned 
away from the imaginary axis. Initially, the second term is suppressed due 
to the small prefactor $T_{K}/\epsilon t$, so that only weak oscillations on top of an exponential 
decay are observed, 
(see Fig.~\ref{fig:QuenchEps1}). At larger times, $T_{K}(t-t_{q})\gtrsim 2\log(\epsilon/T_{K})$, due to the 
twice smaller decay rate the second term dominates, revealing a weaker decay which is overlaid by 
(relatively) strong oscillations.

\begin{figure}
\centering
\includegraphics[width=.9\columnwidth,clip]{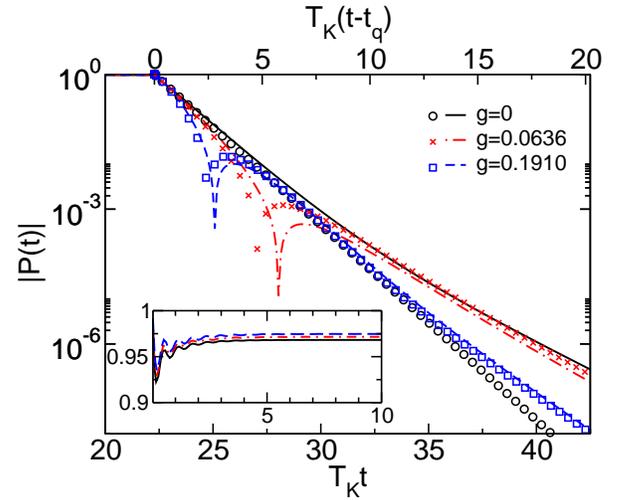}
\caption{(Color online) Time evolution of the spin expectation value $P(t)$ after a quench of the Zeeman splitting 
from $\epsilon/T_K=10$ to zero for different $g$ (lines) obtained within FRG. For comparison 
$|P(t)|$ of the relaxation protocol is shown (symbols). Inset: $P(t)$ before the quench at $T_K t_q=22.5$.} 
\label{fig:QuenchEps2}
\end{figure}

Results for $P(t)$ for the reversed quench from large $\epsilon$ to zero are  
presented in Fig.~\ref{fig:QuenchEps2}. For $g>0$ the behavior of the spin expectation value 
after the quench (lines in  Fig.~\ref{fig:QuenchEps2}) resembles the one found within 
the relaxation protocol in the (partially) coherent regime (symbols  in  Fig.~\ref{fig:QuenchEps2}).
However, there is a substantial difference between the two cases for $g=0$. In the quench protocol
at large $t$ the decay  with the relaxation rate $\Gamma_{0}/2$ dominates which is absent 
in the relaxation protocol. This effect can be understood from the analytical expression for $P(t)$ 
after an $\epsilon$-quench at $g=0$, which is given in Ref.~\onlinecite{Anders06}.  

Non-Markovian memory plays a subdominant role for those 
types of quenches.

\subsubsection{Current in a two-reservoir setup with level position quench}

The developed methods can be easily generalized for the problem of transport through 
a two-lead quantum dot, where the current in the reservoirs can be measured. The following setup is 
considered: at $t=0$ the dot level with large $\epsilon$ is coupled to the reservoirs held at 
different chemical potentials $\mu_{L/R} = \pm V/2$ and the time evolution is performed. 
After relaxation into the 
stationary state, $\epsilon$ is quenched to a value making it resonant with the left lead. 
Due to the finite voltage $V$ applied across  the dot, the model can no longer be mapped onto 
the SBM. We still discuss the time evolution of this model in this paper as it
can partly be understood in terms of the SBM physics described above. 

\begin{figure}
\centering
\includegraphics[width=.9\columnwidth,clip]{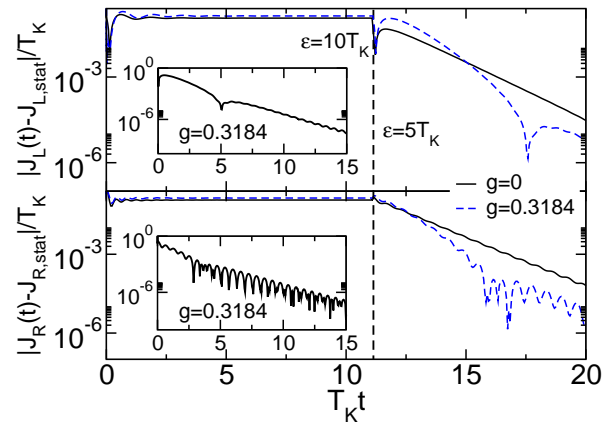}
\caption{(Color online) Time dependence of the left (upper panel) and right (lower panel) currents in a two-lead
setup with a bias voltage $V/T_K=10$ applied symmetrically across the dot. At time $T_K t_q = 11.75$, that is after 
reaching the stationary state with respect to the initial preparation, the level position 
$\epsilon$ is abruptly lowered from $10 T_K$ to that being resonant with the left lead $\epsilon-V/2 =0$.    
Black solid lines show the noninteracting case, while blue dashed lines are FRG data obtained for $g=0.3184$.
Insets: Time evolution of the left and right currents with $\epsilon -V/2 = 0$ initially and no parameter quench 
at a later time.}
\label{fig:current}
\end{figure}

Typical data for the current in the left (upper panel) and right lead (lower panel) in this setup 
are shown in Fig.~\ref{fig:current}. The surprising finding is that despite the large bias $V/T_K = 10$ 
the left lead current only has a slight trace of oscillations with frequency of the 
reservoir bias $V$. In two lead setups with finite bias voltage, the latter usually dominates the 
current (see Refs.~\onlinecite{Andergassen11} and \onlinecite{Kennes12}); in the 
present case this can be inferred from the  right current (dashed blue line in the lower panel). 
Instead, the oscillatory part of the left current is characterized by the SBM frequency 
$\Omega$ appearing for $g>0$; see the dip at $T_K t \approx 17.5$ in the dashed blue 
line of the upper panel of Fig.~\ref{fig:current}. For small $g$, $\Omega$ is proportional 
to $g$ and thus zero for $g=0$ (solid black line in the upper panel).  
This is a clear evidence that 
depending on the precise setup, the largest low-energy scale, in the present case $V$, may not 
serve as a cutoff of the RG flow and is thus not visible in all observables. 

The additional dip in the left lead current right after the quench at $T_K t_q = 11.75$ 
reflects the power-law decay of $P(t)$ in the relaxation protocol at small times given 
in Eq.~\eqref{eq:shorttimesFRG}. During the relaxation process, the tunnel coupling to the left 
lead, which is in resonance with the dot level, is dominating over the right one, that is, fermions 
predominantly hop from the dot level to the left lead and back. The left current can thus be 
approximated as $J_{L}(t)\approx \partial_{t}P(t)$, with $P(t)= 2 \left< d^\dag d \right>(t)-1$, 
and  the time dependence of the current for short times is $\propto t^{g}$. This leads to the first dip 
right after the quench in the upper panel of Fig.~\ref{fig:current}. 

For large times $T_K t \gtrapprox 17$ the right current shows oscillations which do not involve 
transitions through zero; see the dashed blue line in the lower panel of Fig.~\ref{fig:current}
which becomes smooth for $T_K t \gtrapprox 17$. 
Those are absent if one computes the right current with the level being initially 
resonant with the left lead and no additional parameter quench at a later time 
(relaxation protocol); see the lower 
inset. 
In this case dips appear which are associated to zeros of the current. This difference can be 
explained as follows. In the quench protocol in addition to the 
oscillating term with decay rate $\Gamma_{2}^{\ast}/2$, $(T_{K}/\epsilon)e^{-\Gamma_{2}^{*}t/2}\sin(\epsilon t)/(\epsilon t)$ a
 \emph{monotonic} contribution with the same rate 
$\propto e^{-\Gamma_{2}^{*}t/2}$ must appear. Adding those up leads to the observed 
exponential decay with rate $\Gamma_{2}^{\ast}/2$ overlaid by oscillations with frequency 
$\epsilon$. This monotonic term is absent in the relaxation protocol and can thus be attributed to 
the system memory from before the $\epsilon$-quench.

\section{Conclusions}
\label{sec:conclu}

Using two complementary RG methods, we have provided a comprehensive study of the relaxation 
dynamics in the Ohmic SBM. The SBM is considered to be the prototype model 
for a microscopic description of dissipation in open quantum systems.  
We exploit the mapping of the SBM onto the IRLM.
We have studied two basic protocols: a relaxation protocol in which at time $t=0$ the system is prepared 
in a product state and the time evolution is performed with the SBM Hamiltonian as well as 
a quench protocol in which in addition at time $t_q>0$ a parameter of the Hamiltonian is abruptly changed.  
While the FRG was used earlier to study quenches in locally correlated quantum systems we here extended the 
RTRG to this type of nonequilibrium setup.
A comparison of the results obtained using our two approximate methods shows that both provide controlled 
access to the dynamics of the SBM close to the coherent-to-incoherent transition. In the limit of
exponentially large times of the relaxation protocol a comparison to established results derived by other 
approaches is possible and further substantiates this. Crucially, we present new results for intermediate
to long times providing a profound understanding of the details of the relaxation dynamics in all time 
regimes of interest, in particular those relevant for discussing the crossover from coherent to
incoherent behavior.      

Our first central result concerns the classification of the dynamics in the relaxation protocol
with Zeeman splitting $\epsilon=0$ as coherent or incoherent. We have shown that for spin-boson 
couplings $\alpha$ less than 1/2 but close to this value the dynamics should be denoted as partially 
coherent as only a few local minima (and associated zeros) of the spin expectation value 
$P(t)$ can be observed on intermediate times while the asymptotic dynamics is purely monotonic. 
We have confirmed that for $1/2 < \alpha < 1$ the dynamics is monotonic, that 
is incoherent. The main focus of our work is on the role of the non-Markovian memory in the quench 
dynamics. We have identified several situations in which the latter dominates the physics. Most 
prominent are the effects in interaction quenches across the (partially) coherent-to-incoherent 
transition. In this case nonmonotonic behavior can be transferred deep into a regime in which 
the time evolution is performed with coupling constant $\alpha>1/2$ and $P(t)$ might become monotonic 
for times at which $\alpha<1/2$. Both effects are surprising manifestations of non-Markovian memory.  
We have shown that the latter plays a subdominant role in the dynamics resulting out of quenches of 
the Zeeman splitting $\epsilon$.  

We finally left the framework of the SBM and considered a quench in a two-lead IRLM with a finite bias
voltage applied across the dot. If the level is quenched on resonance with one of the leads, SBM physics 
is found in this setup. In addition, non-Markovian memory has a sizable effect in this system.     

For future research in the field of relaxation and quench dynamics in open quantum systems we note
that the RTRG and FRG methods are quite flexible tools to discuss various models, physical quantities
and initial setups. Whereas the RTRG requires a weak system-bath coupling, the FRG needs weak
local interactions in the quantum system. In this respect the two methods are complementary and a
huge parameter regime can be covered. Aside from the local density matrix, other physical observables such as
the particle and heat current for biased situations, spectral densities, and explicitly time-dependent
Hamiltonians can be treated. All correlated initial conditions at time $t=0$ can be studied which can be realized
by the time dynamics out of an uncorrelated setup at some past time $t_0<0$.\cite{Kennes12b} Within the RTRG, this means that 
the system and bath should be decoupled at $t=t_0$, whereas for the FRG the total density matrix at 
$t=t_0$ should not contain any correlations from interactions. At $t=0$, correlations are built up by
the real-time dynamics such that, in the limit $t_0\rightarrow -\infty$, all equilibrium density
matrices containing arbitrary local interactions and system-bath correlations can be used as initial  
condition. Furthermore, by using sequences of different quenches at different times, together with
the possibility to choose a finite value for $t_0$, many more initial conditions even in nonequilibrium
setups can be studied.

An interesting aspect of our approach is the possibility to investigate non-Markovian 
dynamics of open quantum systems and to analyze the relation to concepts of quantum 
non-Markovianity discussed within information theoretic approaches.\cite{Breuer12} Several measures 
have been proposed to characterize the degree to which the dynamics of a given system is non-Markovian.
For weak-coupling models our methods provide controlled access to the reduced density matrix, that is 
all diagonal and nondiagonal matrix elements, from which the proposed measures of non--Markovianity can 
be determined. However, for the specific example of the Ohmic SBM close to the coherent-to-incoherent 
transition (at $\alpha=\frac{1}{2}$), the coupling constant $\alpha$ is of order 1 and we rely on the 
mapping to the IRLM, where a weak-coupling expansion in the parameter $g=1-2\alpha$ is possible. 
As a consequence, we were only able to discuss the spin expectation value of the SBM in the $z$ direction, 
since this observable can be directly related to the local occupation probability of the IRLM. In contrast, 
the spin expectation values in the $x$ or $y$ directions can not be related to the off-diagonal elements of the 
local density matrix of the IRLM but is a rather complicated nonlinear observable involving reservoir 
degrees of freedom. Therefore, for the SBM close to $\alpha=\frac{1}{2}$, it would be interesting  
to investigate the time evolution of the off-diagonal elements of the reduced density matrix using
alternative approaches.

\section*{Acknowledgment}
We thank U.~Weiss for discussions. This work was supported by the DFG via FOR 723.

\appendix

\section{Diagrammatics in Liouville space}
\label{sec:app:diagrams}
We here show how the diagrammatic expansion developed in Ref.~\onlinecite{Schoeller09}
for time-translational invariant systems can be extended to the general case of time-dependent 
Hamiltonians in a straightforward way. We start from a total Hamiltonian of the form 
$H_{\text{tot}}(t)=H_{\text{res}}(t)+H_S(t)+V(t)$, where 
\begin{align}
\nonumber
H_{\text{res}}(t)\,&=\,\sum_\alpha \,[H_\alpha\,+\,\delta\mu_\alpha(t)\,N_{\alpha}]\\
\label{eq:H_res}
&=\sum_{k\nu}[\epsilon_{k\nu}+\delta\mu_\alpha(t)]a^\dagger_{k\nu}a_{k\nu}
\end{align}
describes a set of noninteracting reservoirs with time-dependent chemical potentials
$\mu_\alpha+\delta\mu_\alpha(t)$ ($\alpha$ denotes the reservoir index and $\nu\equiv\alpha n \sigma$
contains in addition the channel index $n$ and the spin index $\sigma$), $H_S(t)$ is any 
time-dependent Hamiltonian operator describing 
the isolated local quantum system, and $V(t)$ is a generic interaction between the local system
and the reservoirs which, following the notation of Ref.~\onlinecite{Schoeller09}, is written in
the compact form ($n=1,2,\dots$)
\begin{equation}
\label{eq:V}
V(t)\,=\,\frac{1}{n!}\,g_{12\dots n}(t)\,:a_1 a_2 \dots a_n: .
\end{equation}
Here, $a_1=\sum_k \delta(\omega-\epsilon_{k\nu}+\mu_\alpha)(a_{k\nu}^\dagger\delta_{\eta +} 
+a_{k\nu}\delta_{\eta -})$ are the reservoir field operators in continuum notation, and 
$1\equiv \eta\nu\omega$ is a multi-index containing the information for creation/annihilation
operators ($\eta=\pm$) and characterizing the reservoir state ($\epsilon_{k\nu}=\omega+\mu_\alpha$ is
the energy of the reservoir state). Implicitly we sum
over all discrete indices $\eta_i$ and $\nu_i$ and integrate over the frequencies $\omega_i$.
$:\dots:$ denotes normal-ordering w.r.t. the reservoir equilibrium distribution and 
$g_{12\dots n}$ is any vertex operator acting only on the
local system characterizing the change of the local state in an interaction process.
For its determination for concrete models respecting the correct sequence of fermionic
field operators we refer to Ref.~\onlinecite{Schoeller09}.

For times $t<t_0$ the local system is assumed to be decoupled from the
reservoirs such that the total initial density matrix has the product form 
\begin{equation}
\label{eq:initial_condition}
\rho_{\text{tot}}(t_0)\,=\,\rho(t_0)\,\rho_{\text{res}}^{\text{eq}} ,
\end{equation}
where $\rho(t_0)$ is any initial density matrix for the local system and 
$\rho_{\text{res}}^{\text{eq}}=\prod_\alpha \rho_{\alpha}^{\text{eq}}$ 
describes each reservoir $\alpha$ in grandcanonical equilibrium
$\rho_{\alpha}^{\text{eq}}=\frac{1}{Z_\alpha}e^{-(H_\alpha-\mu_\alpha N_\alpha)/T_\alpha}$,
characterized by its chemical potential $\mu_\alpha$ and temperature $T_\alpha$ (we set $k_B=\hbar=1$).

To describe the time evolution for times $t>t_0$ we introduce the Liouville operators
$L_{\text{tot}}(t)=[H_{\text{tot}}(t),\cdot]$, $L_{\text{res}}(t)=[H_{\text{res}}(t),\cdot]$,
$L_S(t)=[H_S(t),\cdot]$ and $L_V(t)=[V(t),\cdot]$, such that the time evolution for the
reduced density matrix $\rho(t)=\text{Tr}_{\text{res}}\rho_{\text{tot}}(t)$ of the local system can
be written as ($\mathcal T$ denotes the time-ordering operator)
\begin{align}
\nonumber
\rho(t)\,&=\,\text{Tr}_{\text{res}}\, {\mathcal T} e^{-i\int_{t_0}^t L_{\text{tot}}(t') dt'}\,\rho_{\text{tot}}(t_0)\\
\label{eq:time_evolution}
&=\,\langle {\mathcal T} e^{-i\int_{t_0}^t [L_{\text{res}}(t')+L_S(t')+L_V(t')] dt'}\rangle_{\text{res}}\,\rho(t_0)\,
\end{align}
where $\langle \dots \rangle_{\text{res}}=\text{Tr}_{\text{res}} \dots \rho_{\text{res}}^{\text{eq}}$
denotes the average w.r.t. the equilibrium reservoir distribution. Expanding 
Eq.~(\ref{eq:time_evolution}) in
$L_V$ we find 
\begin{align}
\label{eq:rho_time_propagator}
\rho(t)\,&=\,\Pi(t,t_0)\,\rho(t_0) ,\\
\label{eq:propagator}
\Pi\,&=\,\sum_{m=0}^\infty\,(-i)^m\,\langle\left(\Pi^{(0)}\,(L_V\,\Pi^{(0)})^m\right)\rangle_{\text{res}} ,
\end{align}
where $\Pi(t,t')$ is an effective propagator acting only in Liouville space of the local system, and we 
used a compact matrix notation in time space with the continuum matrix elements
\begin{align}
\label{eq:propagator_0_matrix}
\Pi^{(0)}(t,t')\,&=\,\theta(t-t')\,T\,e^{-i\int_{t'}^t [L_{\text{res}}(\tau)+L_S(\tau)] d \tau } ,\\
\label{eq:Liouvillian_V_matrix}
L_V(t,t')\,&=\,\delta(t-t'-0^+)\,L_V(t) .
\end{align}

To find a diagrammatic representation for $\Pi(t,t')$ and a self-consistent kinetic equation for $\rho(t)$,
we proceed as in Ref.~\onlinecite{Schoeller09} and integrate out the reservoir field operators by
using the representation
\begin{equation}
\label{eq:LV_field_operators}
L_V\,=\,\frac{1}{n!}\,\sigma^{p_1\dots p_n}\,G^{p_1 \dots p_n}_{1\dots n}(t)\,
:A_1^{p_1}\dots A_n^{p_n}: ,
\end{equation}
where $p_i=\pm$ denote the Keldysh indices (over which we sum implicitly), $\sigma^{p_1\dots p_n}$
is a Liouvillian sign operator acting on the local system (only necessary for fermions), and 
$G^{p_1 \dots p_n}_{1\dots n}(t)$ is a Liouvillian vertex operator related to the 
vertex operator $g_{1\dots n}$ (for the precise definitions see Ref.~\onlinecite{Schoeller09}).
$A^p_1$ are reservoir field superoperators acting in Liouville space, defined by 
$A^p_1 b = a_1 b\delta_{p+}+ b a_1\delta_{p-}$ when acting on a reservoir operator $b$.
Inserting Eq.~(\ref{eq:LV_field_operators}) in Eq.~(\ref{eq:propagator}), shifting all field superoperators to the
right by using the identity $A_1^p L_{\text{res}}(t)=(L_{\text{res}}(t)-x_1(t))A_1^p$ with
$x_1(t)=\eta(\omega+\mu_\alpha+\delta\mu_\alpha(t))$, applying Wick's theorem with the
following contraction of field superoperators (the upper/lower case refers to bosons/fermions)
\begin{equation}
\label{eq:contraction_def}
\gamma_{11'}^{pp'}\,=\,
{A_1^p\,A_{1'}^{p'}
  \begin{picture}(-20,11) 
    \put(-22,8){\line(0,1){3}} 
    \put(-22,11){\line(1,0){12}} 
    \put(-10,8){\line(0,1){3}}
  \end{picture}
  \begin{picture}(20,11) 
  \end{picture}
}
\,=\,
\left\{
\begin{array}{cl}
1 \\ p'
\end{array}
\right\}\,
\langle A_1^p\,A_{1'}^{p'}\rangle_{\text{res}} ,
\end{equation}
and using $\text{Tr}_{\text{res}}L_{\text{res}}=0$, we arrive at the diagrammatic representation
\begin{align}
\nonumber
\Pi\,&=\,\Pi_S\,+\,\sum_{m=2}^\infty\,\,\sum_{\text{diagrams}}\,
\frac{(\pm)^{N_p}}{S}\,\left(\prod \gamma\right)\,(-i)^m \\
\label{eq:propagator_diagrams}
&\hspace{1cm}\times  \Pi_S\,(G\,\Pi_S^{X_1})\dots(G\,\Pi_S^{X_{m-1}})\,G\,\Pi_S\,,
\end{align}
with
\begin{align}
\label{eq:Pi_S}
\Pi_S^X(t,t')\,&=\,\theta(t-t')\,T\,e^{-i\int_{t'}^t (L_S+X)(\tau) d\tau} ,\\
\nonumber
G(t,t')\,&\equiv\,(G^{p_1\dots p_n}_{1\dots n})(t,t')\\
\label{eq:G_time}
&=\,\delta(t-t'-0^+)\,G^{p_1\dots p_n}_{1\dots n}(t) ,
\end{align}
and $\Pi_S\equiv\Pi_S^{X=0}$. The quantity $X_i(t)$ appearing in the propagator $\Pi_S^{X_i}$ is 
defined as the sum over all 
$x_j(t)=\eta_j [\omega_j+\mu_{\alpha_j}+\delta\mu_{\alpha_j}(t)]$ from contractions running over
this propagator with the index $j$ stemming from the vertex standing left to this propagator. 
$N_p$ is the number of crossings of fermionic contractions and $S=\prod_i m_i!$ is a symmetry factor
needed for each pair $i$ of vertices connected by $m_i$ equivalent lines (see 
Ref.~\onlinecite{Schoeller09} for more details and the derivation of the diagrammatic rules). 
$\prod\gamma$ stands for the product of all contractions Eq.~(\ref{eq:contraction_def}) and, 
for brevity, the indices of the vertices $G$ have not been indicated in Eq.~(\ref{eq:propagator_diagrams}).
Explicitly, the contraction is given by
\begin{equation}
\label{eq:contraction_explicit}
\gamma_{11'}^{pp'}\,=\,\delta_{1\bar{1}'}
\,p'\,\left\{
\begin{array}{cl}
\eta \\ 1
\end{array}
\right\}\,\rho_\nu(\omega)\,f_\alpha(p'\eta\omega) ,
\end{equation}
where $\bar{1}\equiv-\eta,\nu\omega$ and 
$\delta_{11'}=\delta_{\eta\eta'}\delta_{\nu\nu'}\delta(\omega-\omega')$.
$\rho_\nu(\omega)=\sum_k\delta(\omega-\epsilon_{k\nu}+\mu_\alpha)$ denotes the 
density of states and $f_\alpha(\omega)=1/(e^{\omega/T_\alpha}\mp 1)$ is the 
Bose/Fermi-function of reservoir $\alpha$.

By grouping all diagrams in sequences of irreducible blocks $\Sigma$ (where each propagator has at least
one contraction crossing over it) we find
\begin{equation}
\label{eq:propagator_irr_blocks}
\Pi\,=\,\Pi_S\,\sum_{k=0}^\infty\,(-i\,\Sigma\,\Pi_S)^k ,
\end{equation}
where
\begin{align}
\nonumber
\Sigma\,&=\,i\,\sum_{m=2}^\infty\,\,\sum_{\text{diagrams}}\,
\frac{(\pm)^{N_p}}{S}\,\left(\prod \gamma\right)_{\text{irr}}\,(-i)^m  \\
\label{eq:sigma_diagrams}
&\hspace{1cm}\times  G\,\Pi_S^{X_1}\dots G\,\Pi_S^{X_{m-1}}\,G\,,
\end{align}
denotes the sum over all irreducible diagrams indicated by 
$\left(\prod \gamma\right)_{\text{irr}}$. From (\ref{eq:propagator_irr_blocks}) we
find that the propagator can be determined from the self-consistent equation
\begin{equation}
\label{eq:propagator_self_consistent}
\Pi\,=\,\Pi_S\,(1\,-\,i\,\Sigma\,\Pi) .
\end{equation}
Expanding $\Pi_S$ in $L_S$ we find
\begin{equation}
\label{eq:propagator_S_self_consistent}
\Pi_S\,=\,\theta\,(1\,-\,i\,L_S\,\Pi_S) , 
\end{equation}
where $L_S(t,t')=L_S(t)\delta(t-t'-0^+)$ and $\theta(t,t')=\theta(t-t')$.
Inserting this equation in Eq.~(\ref{eq:propagator_self_consistent}) we obtain
\begin{align}
\label{eq:prop_1}
\Pi\,&=\,\theta\,(1\,-\,i\,L\,\Pi)\\
\label{eq:prop_2}
&=\,\theta\,\sum_{k=0}^\infty\,(-i\,L\,\theta)^k ,
\end{align}
where we have introduced the effective Liouvillian
\begin{equation}
\label{eq:effective_Liouvillian}
L\,=\,L_S\,+\,\Sigma .
\end{equation}

A self-consistent kinetic equation for $\rho(t)$ can be derived by taking the
time derivative of Eqs.~(\ref{eq:rho_time_propagator}) and (\ref{eq:prop_1}). We find
\begin{equation}
\label{eq:partial_t_prop}
\partial_t\Pi\,=\,\delta\,-\,i\,L\,\Pi ,
\end{equation}
where $(\partial_t\Pi)(t,t')=\frac{\partial}{\partial t}\Pi(t,t')$
and $\delta(t,t')=\delta(t-t'-0^+)$. Inserting this relation in 
$\dot{\rho}(t)=(\partial_t\Pi)(t,t_0)\rho(t_0)$ leads to the kinetic equation 
\begin{equation}
\label{eq:kinetic_equation}
i\,\dot{\rho}(t)\,=\,\int_{t_0}^t L(t,t')\,\rho(t') dt'.
\end{equation}

We note that the compact matrix notation in time space allows to write
all equations straightforwardly in Fourier space by using the Dirac
notation $\langle t|E\rangle=\frac{1}{\sqrt{2\pi}}e^{-iEt}$. This leads
to 
\begin{align}
\label{eq:L_S_fourier}
L_S(E,E')\,&=\,\frac{1}{2\pi}\,\int e^{i(E-E')t}\,L_S(t) dt\\
\label{eq:A_fourier}
A(E,E')\,&=\,\frac{1}{2\pi}\,\iint e^{iEt-iE't'}\,A(t,t') dt dt' ,\\
\label{eq:delta_fourier}
\delta(E,E')\,&=\,\frac{1}{2\pi}\,\int e^{i(E-E')t} dt =  \delta(E-E') ,\\
\nonumber
\theta(E,E')\,&=\,\frac{1}{2\pi}\,\iint_{t>t'} e^{iEt-iE't'} dt dt'\\
\label{eq:theta_fourier}
&=\,\frac{i}{E}\,\delta(E-E') ,
\end{align}
where $A\equiv\Pi,\Pi_S,L,\Sigma$.
For $\rho(E)$ we use the special definition
\begin{equation}
\label{eq:rho_fourier}
\rho(E)\,=\,\int\,dt\,e^{iE(t-t_0)}\,\rho(t)\,\theta(t-t_0) ,
\end{equation}
such that Eqs.~(\ref{eq:rho_time_propagator}) and (\ref{eq:kinetic_equation}) read in Fourier space
\begin{align}
\label{eq:rho_fourier_propagator}
\rho(E)\,&=\,\sqrt{2\pi}\,e^{-iEt_0}\,\Pi(E,t_0)\,\rho(t_0) ,\\
E\,\rho(E)\,&=\,i\,\rho(t_0) \nonumber \\
\label{eq:kinetic_equation_fourier}
&+\,\int L(E,E')\,e^{-i(E-E')t_0}\,\rho(E') dE'.
\end{align}
For the special case of a time-translational invariant Hamiltonian all objects
$A(t,t')=A(t-t')$ depend only on the relative time difference and $L_S$ is 
independent of time. This gives
\begin{align}
\label{eq:L_S_fourier_relative_time}
L_S(E,E')\,&=\,\delta(E-E')\,L_S\\
\nonumber
A(E,E')\,&=\,\delta(E-E')\,\int_0^\infty e^{iEt}\,A(t) dt  \\
\label{eq:A_fourier_relative_time}
&\equiv\,\delta(E-E')\,A(E)\\
\label{eq:pi(E,t_0)}
\Pi(E,t_0)\,&=\,\frac{1}{\sqrt{2\pi}}\,e^{iEt_0}\,\Pi(E)
\end{align}
Inserting Eqs.~(\ref{eq:delta_fourier}), (\ref{eq:theta_fourier}), 
(\ref{eq:L_S_fourier_relative_time}), and (\ref{eq:pi(E,t_0)}) in 
(\ref{eq:propagator_S_self_consistent}), (\ref{eq:prop_1}) and
(\ref{eq:rho_fourier_propagator}) we obtain
\begin{align}
\label{eq:pi_S_relative_time}
\Pi_S(E)\,&=\,\frac{i}{E\,-\,L_S} ,\\
\label{eq:pi_relative_time}
\Pi(E)\,&=\,\frac{i}{E\,-\,L(E)} ,\\
\label{eq:rho_relative_time}
\rho(E)\,&=\,\Pi(E)\,\rho(t_0) .
\end{align}
For the diagrammatic representation of $\Sigma(E)$ we obtain
from Eq.~(\ref{eq:sigma_diagrams}) the result
\begin{align}
\nonumber
\Sigma(E)\,&=\,\sum_{m=2}^\infty\,\,\sum_{\text{diagrams}}\,
\frac{(\pm)^{N_p}}{S}\,\left(\prod \gamma\right)_{\text{irr}} \\
\label{eq:sigma_diagrams_relative_time}
&\hspace{-1cm}\times G\,R_S(E+X_1)\dots G\,R_S(E+X_{m-1})\,G\,,
\end{align}
with the resolvent $R_S(E)=-i\Pi_S(E)=1/(E-L_S)$. Conveniently,
the frequency integration over the variables $\bar{\omega}=\eta\omega$
appearing in the quantities $X_i$ are performed by closing the integration
contour in the upper half of the complex plane where, except for the
density of states and the Bose/Fermi-function appearing in the 
contraction Eq.~(\ref{eq:contraction_explicit}), the integrand is analytic.

\section{Quench dynamics}
\label{sec:app:quench_dynamics}
If, at a certain quench time $t=t_q$, the properties of the Hamiltonian are
discontinuously changed, it is convenient to define propagators and effective
Liouville operators corresponding
to the time evolution after and before the quench as well as to the memory by 
\begin{align}
\label{eq:A_f}
A_f(t,t')\,&=\,A(t,t')\,\theta(t_q-t') ,\\
\label{eq:A_i}
A_i(t,t')\,&=\,\theta(t_q-t)\,A(t,t') ,\\
\label{eq:A_fi}
A_{fi}(t,t')\,&=\,\theta(t-t_q)\,A(t,t')\,\theta(t_q-t') ,
\end{align}
with $A\equiv\Pi,\Pi_S,\Sigma,L,L_S$. Since $L_{S,fi}=0$ we get $L_{fi}=\Sigma_{fi}$.
Due to the form (\ref{eq:prop_2}) of the propagator we obtain the central equation
\begin{align}
\label{eq:prop_quench}
\Pi_{fi}(t,t')\,=\,\Pi_f(t,t_q)\,\Pi_i(t_q,t')\,-\,i\,(\Pi_f\,\Sigma_{fi}\,\Pi_i)(t,t')\,,
\end{align}
which holds in the generic case even if the Hamiltonian is time-dependent before and
after the quench. The relation between the density matrix $\rho_f(t)=\rho(t)\theta(t-t_q)$
after the quench and the one at the initial time $t_0<t_q$ before the quench 
(where the local system and the reservoirs are assumed to be decoupled) can be calculated from
\begin{equation}
\label{eq:rho_after_quench}
\rho_f(t)\,=\,\Pi_{fi}(t,t_0)\,\rho(t_0) .
\end{equation}

For the special case of a time-independent Hamiltonian before and
after the quench (which we discuss in this paper), the propagators $\Pi_{f/i}$ after
and before the quench depend only on the relative time difference. Introducing the special 
Fourier transformation for the memory parts (where $t_q$ is used as
reference time and the prefactor $\frac{1}{2\pi}$ is omitted)
\begin{equation}
\label{eq:fi_fourier}
A_{fi}(E,E')\,=\,\iint e^{iE(t-t_q)-iE'(t'-t_q)}\,A_{fi}(t,t') dt dt',
\end{equation}
and using Eq.~(\ref{eq:A_fourier_relative_time}) we obtain for Eq.~(\ref{eq:prop_quench}) 
in Fourier space 
\begin{equation}
\label{eq:prop_fi_fourier}
\Pi_{fi}(E,E')\,=\,\Pi_f(E)\,(1\,-\,i\,\Sigma_{fi}(E,E'))\,\Pi_i(E')\,,
\end{equation}
with
\begin{align}
\label{eq:prop_f_fourier}
\Pi_f(E)\,&=\,\frac{i}{E\,-\,L_f(E)} ,\\
\label{eq:prop_i_fourier}
\Pi_i(E')\,&=\,\frac{i}{E'\,-\,L_i(E')} .
\end{align}
The diagrammatic representation of this formula is illustrated in Fig.~\ref{fig:diagrams}(a).

The diagrammatic expansion for $-i\Sigma(E,E')$ can be obtained from Eq.~(\ref{eq:sigma_diagrams}) as
\begin{align}
\nonumber
-i\,\Sigma_{fi}(E,E')\,&=\,\iint e^{iE(t-t_q)-iE'(t'-t_q)} \\
\nonumber
&\hspace{-2cm}\times \sum_{m,m'=1}^\infty\,\,\sum_{\text{diagrams}}\,
\frac{(\pm)^{N_p}}{S}\,\left(\prod \gamma\right)_{\text{irr}}\,(-i)^{m+m'} \\
\nonumber
&\hspace{-2.0cm}\times (G_f\,\Pi_{S,f}^{X_1^f}\dots G_f\,\Pi_{S,f}^{X_m^f})(t,t_q) \\
\label{eq:sigma_fi_diagrams_zw}
&\hspace{-2.0cm}\times (\Pi_{S,i}^{X_m^i}\,G_i\,\Pi_{S,i}^{X_1^{\prime,i}}
\dots G_i\,\Pi_{S,i}^{X_{m'-1}^{\prime,i}}\,G_i)(t_q,t') dt dt' ,
\end{align}
where $X_k^f$, $k=1,\dots,m$, contain the chemical potentials after the quench, whereas the
variables $X_m^i$ and $X_{k'}^{\prime,i}$, $k'=1,\dots,m'$ involve the chemical potentials before the quench.
Performing the time integrals and using that $G_{f/i}$ is independent of time and
$\Pi_{S,f/i}(t,t')=\Pi_{S,f/i}(t-t')$ depends only on the relative time difference, gives a product in 
Fourier space for the two expressions left and right to the quench with the result
\begin{align}
\nonumber
-i\,\Sigma_{fi}(E,E')\,&=\,\sum_{m,m'=1}^\infty\,\,\sum_{\text{diagrams}}\,
\frac{(\pm)^{N_p}}{S}\,\left(\prod \gamma\right)_{\text{irr}} \\
\nonumber
&\hspace{-2.2cm}\times G_f\,R_S^f(E+X_1^f)\dots G_f\,R_S^f(E+X_m^f)\,R_S^i(E'+X_m^i) \\
\label{eq:sigma_fi_diagrams}
&\hspace{-2.2cm}\times G_i\,R_S^i(E'+X_1^{\prime,i})
\dots G_i\,R_S^i(E'+X_{m'-1}^{\prime,i})\,G_i\,,
\end{align}
with the resolvents $R_S^{f/i}(E)=-i\Pi_{S,f/i}(E)=1/(E-L_S^{f/i})$.
As a result, we get the usual diagrammatic rules with the difference that the quench time
has to be introduced in one propagator, all resolvents left (right) to the quench contain
the Fourier variable $E$ ($E'$), and all vertices, Liouvillians, and chemical potentials
appearing left (right) to
the quench are associated with the ones after (before) the quench. Furthermore, 
following Ref.~\onlinecite{Pletyukhov12}, it is possible to sum over all contractions 
which do not cross over the quench, such that the full effective vertices $G_{f/i}(E)$ and the full 
resolvents $R_{f/i}(E)=1/[E-L_{f/i}(E)]$ appear left/right to the quench with the result
\begin{align}
\nonumber
-i\,\Sigma_{fi}(E,E')\,&=\,\sum_{m,m'=1}^\infty\,\,\sum_{\text{diagrams}}\,
\frac{(\pm)^{N_p}}{S}\,\left(\prod \gamma\right)_{\text{quench}} \\
\nonumber
&\hspace{-2cm}\times G_f(E)\,R_f(E+X_1^f)\dots G_f(E+X_{m-1}^f)\,R_f(E+X_m^f) \\
\nonumber
&\hspace{-2cm}\times R_i(E'+X_m^i)\,G_i(E'+X_m^i)\,R_i(E'+X_1^{\prime,i})\dots\\
\label{eq:sigma_fi_diagrams_effective}
&\hspace{-2cm}\dots G_i(E'+X_1^{\prime,i})\dots R_i(E'+X_{m'-1}^{\prime,i})\,
G_i(E'+X_{m'-1}^{\prime,i})\,.
\end{align}
In this form all contractions have to cross over the quench, indicated by
$\left(\prod \gamma\right)_{\text{quench}}$, leading automatically to a 
connected diagram. Examples of diagrams are shown in Fig.~\ref{fig:diagrams}(b).

\section{RTRG for IRLM}
\label{sec:app:rtrg_irlm}

Here we present the results of the RTRG method for the IRLM, as they have been derived in
Ref.~\onlinecite{Andergassen11} by using the Matsubara cutoff scheme or alternatively 
in Ref.~\onlinecite{Schoeller13} by the $E$-RTRG method. We consider the special
case of one single reservoir at zero temperature with chemical potential $\mu=0$ and 
consider the particle-hole symmetric case $\epsilon=0$. 

The quantities $Z'(E)$ and $\tilde{L}_\Delta(E)$ defined in (\ref{eq:Zprime_tildeL_def})
are $4\times4$ matrices in Liouvillian space in the basis $(00,11,10,01)$, where $0/1$ 
denote the unoccupied/occupied state of the local state.
Writing each $4\times 4$ matrix in terms of four $2\times 2$ blocks, $Z'(E)$
and $\tilde{L}_\Delta(E)$ can be written as
\begin{align}
\label{eq:Zprime_matrix}
Z'(E)\,&=\,
\left(
\begin{array}{c|c}
\mathbbm{1}  & 0  \\
\hline
 0 &  Z(E)\mathbbm{1}  \\
\end{array}
\right)
 ,\\
\label{eq:tilde_L_matrix}
\tilde{L}_\Delta(E)\,&=\,-i\,
\left(\begin{array}{c|c}
\Gamma_1(E)\tau_- & 0\\
\hline
0 & \frac{1}{2}\Gamma_2(E)\mathbbm{1}\\
\end{array}\right)
,
\end{align}
where $\tau_\pm=\frac{1}{2}(\mathbbm{1}\pm\sigma_x)$, and the functions $Z(E)=Z(-E^*)^*$ and 
$\Gamma_{1/2}(E)=\Gamma_{1/2}(-E^*)^*$ fulfill the RG equations
\begin{align}
\label{eq:RG_Z}
\frac{\partial}{\partial E}Z(E)\,&=\,U^2\,\frac{Z(E)}{E\,+\,i\Gamma_2(E)/2} ,\\
\label{eq:RG_Gamma_1}
\frac{\partial}{\partial E}\Gamma_1(E)\,&=\,-g\,\frac{\Gamma_1(E)}{E\,+\,i\Gamma_2(E)/2} ,\\
\label{eq:RG_Gamma_2}
\frac{\partial}{\partial E}\Gamma_2(E)\,&=\,-g\,\frac{\Gamma_1(E)}{E\,+\,i\Gamma_1(E)} ,
\end{align}
with $g=2U-U^2$. The initial conditions at $E=i\omega_c$ are given by 
$Z=1$ and $\Gamma_1=\Gamma_2=\Gamma_0$.

With Eq.~(\ref{eq:tilde_L_matrix}), the resolvent $\tilde{R}_\Delta(E)=1/(E-\tilde{L}_\Delta(E))$ 
defined in Eq.~(\ref{eq:tilde_prop}) can be decomposed as
\begin{align}
\nonumber
\tilde{R}_\Delta(E)\,&=\,
\frac{1}{E}\,\left(\begin{array}{c|c}\tau_+ & 0 \\ \hline 0 & 0 \\ \end{array}\right)\,+\,
\frac{1}{E+i\Gamma_1(E)}\,\left(\begin{array}{c|c}\tau_- & 0 \\ \hline 0 & 0 \\ \end{array}\right)\\
\label{eq:tilde_R_matrix}
&\hspace{1cm}
+\,\frac{1}{E+i\Gamma_2(E)/2}\,\left(\begin{array}{c|c}0 & 0 \\ \hline 0 & \mathbbm{1} \\ \end{array}\right)
\end{align}

The effective vertices $\tilde{G}_1(E)=Z'(E)G_1(E)$ and $\tilde{G}_{12}(E)=Z'(E)G_{12}(E)$,
where $G_{1}=\sum_{p}G_{1}^{p}$ and $G_{12}=\sum_{pp'}G_{12}^{pp'}$ denote the vertices
averaged over the Keldysh indices, contain only the index $1\equiv\eta=\pm$ characterizing
creation/annihilation reservoir field operators (note that we consider the single reservoir
case without spin). The effective vertices are explicitly given by
\begin{align}
\nonumber
\tilde{G}_{+}(E) \,&=\, \sqrt{\frac{Z(E)\Gamma_{1}(E)}{2\pi}} \\
\label{eq:G_+}
&\times 
\left(\begin{array}{c|c} 
0 & \begin{array}{cc} 1/Z(E) & 0 \\ -1/Z(E) & 0 \\ \end{array}\\
\hline
\begin{array}{cc} 0 & 0 \\ 1-i\pi U & 1+i\pi U \\ \end{array} & 0 \\
\end{array}\right)
 ,\\
\nonumber
\tilde{G}_{-}(E) \,&=\, \sqrt{\frac{Z(E)\Gamma_{1}(E)}{2\pi}} \\
\label{eq:G_-}
&\times
\left(\begin{array}{c|c} 
0 & \begin{array}{cc} 0 & -1/Z(E) \\ 0 & 1/Z(E) \\ \end{array}\\
\hline
\begin{array}{cc} 1+i\pi U & 1-i\pi U \\ 0 & 0 \\ \end{array} & 0 \\
\end{array}\right)
,\\
\label{eq:G_+-}
\tilde{G}_{+-}(E) \,&=\, -\tilde{G}_{-+}(E) \,=\,
\left(\begin{array}{c|c}
0 & 0 \\
\hline
0 & U\sigma_z \\
\end{array}\right) 
,
\end{align}
and $\tilde{G}_{++}=\tilde{G}_{--}=0$.

\section{Branching point position}
\label{sec:app:rrc}
Here we show how to derive the improved formula (\ref{eq:gamma_2*}) for the
position of the branching point $z_0=-i\Gamma_2^*/2$ of the resolvent $R_1(E)=1/(E+i\Gamma_1(E))$,
which is at the same time the pole of the resolvent $R_2(E)=1/(E+i\Gamma_2(E)/2)$, i.e. fulfills
the equation $z_0=-i\Gamma_2(z_0)/2$ or $\Gamma_2(z_0)=\Gamma_2^*$.
We start from the RG equations (\ref{eq:rgflow}) for $E=z_0+i\Lambda$, with $\Lambda>0$, i.e.
the regime of the imaginary axis above the branch cut of $R_1(E)$. By comparison with the numerical 
solution we find that a very good approximation consists in replacing 
$\Gamma_{1/2}(E)\rightarrow \Gamma_2^*$ on the r.h.s. of the RG equations.
This leads to the solution Eq.~(\ref{eq:gam1_solution_all_en}) for $\Gamma_1(E)$, which, when
inserted in the RG equation for $\Gamma_2(E)$ and using $E=-i\Gamma_2^*/2+i\Lambda$
gives the following differential equation for the determination of $\Gamma_2$
\begin{equation}
\label{eq:gam2_diff_equation}
\frac{d \Gamma_2}{d \Lambda}\,=\,-g\,\frac{T_K}{\Lambda+\Gamma_2^*/2}\,
\left(\frac{T_K}{\Lambda}\right)^g .
\end{equation}
This equation has to be solved with the boundary conditions 
\begin{align}
\label{eq:boundary_large}
\Gamma_2\,\,\stackrel{\Lambda\rightarrow\infty}{\longrightarrow}\,\,
&T_K\,\left(\frac{T_K}{\Lambda}\right)^{g} \\
\label{eq:boundary_small}
\Gamma_2\,\,\stackrel{\Lambda\rightarrow 0}{\longrightarrow}\,\,&\Gamma_2^*
\end{align}
Taking another derivative with respect to $\Lambda$ and defining the variable
$x=-\Lambda/(\Gamma_2^*/2)$, we obtain the following special case of the hypergeometic
differential equation
\begin{equation}
\label{eq:hypo_2}
x(1-x)\frac{d^2\Gamma_2}{dx^2}+(g-(1+g)x)\frac{d\Gamma_2}{dx}=0\,.
\end{equation}
Denoting the hypergeometric function by $F(a,b,c,x)$, this equation has the following 
two elementary solutions 
\begin{align}
\label{eq:elementary_solutions}
F(0,g,g,x)\,=\,1 \quad (-x)^{1-g}\,F(1-g,1,2-g,x)\,\,.
\end{align}
Since the hypergeometric function is analytic for $|x|<1$, the second elementary
solution has a branch cut from the power law $(-x)^{1-g}$, which is chosen such 
that the branch cut for $\Gamma_2(E)$ lies on the negative imaginary axis. The
solution for $|x|<1$ can then be written as
\begin{equation}
\label{eq:linear_combination}
\Gamma_2\,=\,\Gamma_2^*\,+\,\lambda\,(-x)^{1-g}\,F(1-g,1,2-g,x)\,\,,
\end{equation}
where we have used the boundary condition Eq.~(\ref{eq:boundary_small}) that $\Gamma_2=\Gamma_2^*$
for $x\rightarrow 0$. The coefficient $\lambda$ can be determined by taking the
derivative w.r.t. $x$ and comparing with the differential equation (\ref{eq:gam2_diff_equation})
for $x\rightarrow 0$. Using $F(a,b,c,0)=1$ we obtain from Eq.~(\ref{eq:linear_combination})
\begin{equation}
\nonumber
\frac{d\Gamma_2}{d\Lambda}=-\frac{2}{\Gamma_2^*}\frac{d\Gamma_2}{dx}
\,\,\stackrel{x\rightarrow 0}{\longrightarrow}\,\,\frac{2}{\Gamma_2^*}\lambda (1-g)
\left(\frac{\Gamma_2^*/2}{\Lambda}\right)^g\,,
\end{equation}
which, when compared with Eq.~(\ref{eq:gam2_diff_equation}), gives 
$\lambda=-\frac{g}{1-g}T_K(\frac{T_K}{\Gamma_2^*/2})^g$, i.e. the solution for $|x|<1$ reads
\begin{align}
\nonumber
\Gamma_2\,&=\,\Gamma_2^*\,-\,\\
\label{eq:gam2_x<1}
&-\frac{g}{1-g}T_K\left(\frac{T_K}{\Gamma_2^*/2}\right)^g
(-x)^{1-g}F(1-g,1,2-g,x)\,.
\end{align}
To find the analytical continuation to the regime $|x|>1$ we use the relation
\begin{widetext}
\begin{align}
\nonumber
F(1-g,1,2-g,x)\,&=\,\frac{\Gamma(2-g)\Gamma(g)}{\Gamma(1)\Gamma(1)}(-x)^{g-1}F(1-g,0,1-g,\frac{1}{x})
+\frac{\Gamma(2-g)\Gamma(-g)}{\Gamma(1-g)\Gamma(1-g)}(-x)^{-1}F \left(1,g,1+g,\frac{1}{x} \right)\\
\label{eq:F_x>1}
&=\,(1-g)\,\frac{\pi}{\sin(\pi g)}\,(-x)^{g-1}\,+\,\frac{1-g}{g}\,\frac{1}{x}\,F \left(1,g,1+g,\frac{1}{x} 
\right) ,
\end{align}
where we have used $\Gamma(2-g)/\Gamma(1-g)=(1-g)$, $\Gamma(1-g)/\Gamma(-g)=-g$ and
$\Gamma(2-g)\Gamma(g)=(1-g)\Gamma(1-g)\Gamma(g)=(1-g)\frac{\pi}{\sin(\pi g)}$ in the last 
step. Inserting Eq.~(\ref{eq:F_x>1}) in Eq.~(\ref{eq:gam2_x<1}) gives the following solution for $|x|>1$
\begin{equation}
\label{eq:gam2_x>1}
\Gamma_2\,=\,\Gamma_2^*
\,-\,\frac{\pi g}{\sin(\pi g)}\,T_K\,\left(\frac{T_K}{\Gamma_2^*/2}\right)^g
\,+\,T_K\,\left(\frac{T_K}{\Lambda}\right)^g\,F \left( 1,g,1+g,\frac{1}{x} \right) .
\end{equation}
\end{widetext}
Comparing this solution with the asymptotic boundary condition Eq.~(\ref{eq:boundary_large}), we
find that the first two terms on the r.h.s. of Eq.~(\ref{eq:gam2_x>1}) have to cancel each other,
leading to the result Eq.~(\ref{eq:gamma_2*}) for $\Gamma_2^*$
\begin{equation}
\label{eq:gamma_2*_app}
\frac{\Gamma_2^*}{2}\,\approx\,\,T_K\,\left(\frac{\pi g}{2 \sin(\pi g)}\right)^\frac{1}{1+g}\,.
\end{equation}

\section{Interaction quenches}
\label{sec:app:kern}

In this Appendix we present the detailed calculation of the second term 
of Eq.~\eqref{eq:PquenchI} for the interaction quench between the coherent 
and the incoherent regimes. To compute the memory contribution to $P(t)$ we start from 
Eq.~\eqref{eq:PquenchF} and restrict ourselves to small couplings $|g_{i/f}|\ll 1$ such
that $g_{i/f}\approx 2 U_{i/f}$.
We use the notation $a=i/f$, $t_i=t_q$ and $t_f=t-t_q$, and calculate the functions 
$F_\Lambda^{a}(t_{a})$ 
in leading order in $g_a$ by dropping all contributions from the resolvents of order ${\mathcal O}(g_a)$. 
Therefore, we replace $\Gamma_2^{a}(E)\rightarrow\Gamma_2^{*a}$ in the resolvent $R_2^{a}(E)$ and 
approximate the $Z$-factors $Z_a(E)\approx 1$. The latter follows from the solution of the RG equation
\eqref{eq:RG_Z}, which in leading order reads as
\begin{equation}
\label{eq:Z_factor}
Z_{a}(E)\,\approx\,\left(\frac{-iE+\frac{1}{2}\Gamma_2^{*a}}{\omega_c}\right)^{U_{a}^2} .
\end{equation}
For $|E|\sim T_K^{a}$ this gives $Z_{a}\sim (T_K^{a}/\omega_c)^{U_{a}^2} \sim 1$ for small $|U_{a}|$.
With these approximations, Eq.~\eqref{eq:PquenchF} reads as
\begin{equation}
F^{a}_{\Lambda}(t_{a}) \approx \frac{1}{2}\sum_{\sigma=\pm}\!\!\int\!\! 
\frac{\sqrt{\Gamma_{1}^{a}(E)}}{E \!+\! i\Gamma_{1}^{a}(E)}
\frac{e^{-iEt_{a}} }{E \!+\! i(\Lambda\!+\!\frac{1}{2}\Gamma_{2}^{*a})\!-\!\sigma 0^+}\frac{dE}{2\pi},
\label{juhu}
\end{equation}
where we added a small imaginary part $i\sigma 0^+$ to $\Lambda$ and have taken the average
of $\sigma = \pm$ in order to define the integrand on the branch-cut of the first resolvent.
Closing the integration contour in the lower half-plane, 
there are two pole contributions at $E=\pm\Omega_{a}-i\Gamma_1^{*a}$ of the first resolvent
(for $g_{a}>0$),
one pole contribution at $E=-i(\Lambda+\frac{1}{2}\Gamma_2^{*a})+\sigma 0^+$ of the second resolvent,
and a branch cut contribution starting at $E=-i\frac{1}{2}\Gamma_2^{*a}$ from the function $\Gamma_1^{a}(E)$.
Neglecting terms of ${\mathcal O}(g_a)$, we consider only the contribution of ${\mathcal O}(1)$ of 
the branch cut. This part is relevant in the regime of intermediate times $\Gamma_1^{*a}t_a\sim 1/g_a$
and stems from the Lorentzian form of the jump of the resolvent $R_1^a(E)$ across the branch cut centered at 
$E\sim -i\Gamma_1^{*a}$ [see the discussion around Eq.~\eqref{eq:lorentz_peak}]. Therefore,
we approximate this peak by a true $\delta$-function such that the branch-cut contribution
is approximately given by a pole contribution at $E=-i\Gamma_1^{*a}$ with 
$R_1^a(E)\approx -\text{sign}(g_a)/(E+i\Gamma_1^{*a})$ close to this pole according to
Eq.~\eqref{eq:lorentz_peak}.

Using Eq.~\eqref{eq:gam1_solution_all_en} for $\Gamma_1^{a}(E)$ and 
neglecting terms ${\mathcal O}(g_{a})$, we can split the various pole contributions as
\begin{equation}
F^{a}_{\Lambda}(t_{a}) \approx F^{a}_{\Lambda,1}(t_{a}) + F^{a}_{\Lambda,2}(t_{a})\,,
\end{equation}
with 
\begin{align}
\nonumber
F^{a}_{\Lambda,1}(t_{a}) &=  
\left(2\cos(\Omega_{a}t_{a})\,\theta(g_{a})-\text{sign}(g_a)\right)\,e^{-\Gamma_1^{*a}t_{a}} \\
\label{eq:Ft1}
&\hspace{1cm}
\times \,\text{Re}\,
\frac{\sqrt{\Gamma_{1}^{*a}}}{\Gamma_{1}^{*a}-\frac{1}{2}\Gamma_{2}^{*a}-\Lambda+i0^+}\,,\\
F^{a}_{\Lambda,2}(t_a) &= -\,\text{Re}\,
\frac{%
\sqrt{T_{K}^{a}}\left[\frac{T_{K}^{a}}{\Lambda}\right]^{g_{a}/2}
e^{-(\Lambda+\frac{1}{2}\Gamma_{2}^{*a})t_a}}
{T_{K}^{a}\left[\frac{T_{K}^{a}}{\Lambda}\right]^{g_{a}}\!\!
 -\frac{1}{2}\Gamma_{2}^{*a}-\Lambda+i0^+}.
\label{eq:Ft2}
\end{align}
Note that when considering the product $F^f_{\Lambda}(t_f)F^i_{\Lambda}(t_i)$ entering 
the second term for $P(t)$ in Eq.~(\ref{eq:PquenchI}), there is no divergence for the 
subsequent integral over $\Lambda$ close to $\Lambda\sim \Gamma_1^{*a}-\frac{1}{2}\Gamma_2^{*a}$
since the principal values of the two resolvents are centered at different positions.
The combination $F^f_{\Lambda,1}(t_f)F^{i}_{\Lambda,1}(t_{i})$ will be
neglected in the following since it has the strongest decay 
$\sim e^{-\Gamma_1^{*f}t_f} e^{-\Gamma_1^{*i}t_i}$. 
For the case $g_a=-g_{\bar{a}}<0$ (with $\bar{a}=f/i$ for $a=i/f$),
the combination $F^a_{\Lambda,2}(t_a)F^{\bar{a}}_{\Lambda,1}(t_{\bar{a}})$ 
contains an exponentially decaying factor $e^{-\Lambda t_a}$
for the integration over $\Lambda$ in (\ref{eq:PquenchI}). This restricts the
integration range to $\Lambda\lesssim 1/t_a\ll\Gamma_1^{*a}<\Gamma_1^{*,\bar{a}}$,
where we have used that $\Gamma_1^{*a}\approx 2^{g_a}T_K^a$ is larger for a positive interaction
than for a negative one. Therefore, we can neglect $-\Lambda+i0^+$ in all denominators of the 
resolvents occurring in Eqs.~(\ref{eq:Ft1}) and 
(\ref{eq:Ft2}). Furthermore, we can approximate $(T_K^a/\Lambda)^{g_a}\approx (T_K^a t_a)^{g_a}$
in Eq.~\eqref{eq:Ft2}.
The same holds for the combination $F^f_{\Lambda,2}(t_f)F^{i}_{\Lambda,2}(t_i)$, except
that an exponential factor $e^{-\Lambda(t_i+t_f)}=e^{-\Lambda t}$ occurs such that we 
get $(T_K^a/\Lambda)^{g_a}\approx (T_K^a t)^{g_a}$ in Eq.~\eqref{eq:Ft2}.
With these approximations we can easily calculate the final integral over $\Lambda$
to get the second term of Eq.~\eqref{eq:PquenchI} for the two quench protocols.

\paragraph{In the coherent to incoherent quench} the coupling 
before $t_{q}$ is positive and has the same absolute value as the one afterwards: 
$g_{i}=-g_{f}=g>0$. The combinations $F^f_{\Lambda,2}(t_f)F^{i}_{\Lambda,1}(t_i)$ 
and $F^f_{\Lambda,2}(t_f)F^{i}_{\Lambda,2}(t_i)$ are both purely decaying
after the quench with decay rate $\frac{1}{2}\Gamma_2^{*f}$. However, since 
$F^i_{\Lambda,1}(t_i)\sim e^{-\Gamma_1^{*i}t_i}$ and 
$F^i_{\Lambda,2}(t_i)\sim e^{-\frac{1}{2}\Gamma_2^{*i}t_i}$, the combination 
$F^f_{\Lambda,2}(t_f)F^{i}_{\Lambda,2}(t_i)$ will dominate. Together with $-g_i=-g$,
$\Gamma_2^{*i}\approx 2^g T_K^i$ and $\Gamma_2^{*f}\approx 2^{-g}T_K^f$, this leads to the result
\begin{align}
\nonumber
-g_i\,\int_{0}^{\infty} F_{\Lambda,2}^{f}(t-t_q)F_{\Lambda,2}^{i}(t_q) d\Lambda &\approx
-g\frac{e^{-\frac{1}{2}\Gamma_{2}^{*i} t_{q}-\frac{1}{2}\Gamma_{2}^{*f} (t-t_{q})}}{t}\\
&\hspace{-4cm}\times
\frac{1}{\sqrt{T_K^{i}T_K^{f}}}
\,
\frac{\left(T_{K}^{i}t/2\right)^{g/2}}{\left(T_{K}^{i}t/2\right)^{g}-\frac{1}{2}}
\,
\frac{\left(T_{K}^{f}t/2\right)^{-g/2}}{\left(T_{K}^{f}t/2\right)^{-g}-\frac{1}{2}}.
\label{eq:F2F2:c-ic}
\end{align}
Introducing the  function $S^{\pm}_{x}$ defined in Eq.~\eqref{eq:Sdef} in the main text, 
we obtain the second term of Eq.~\eqref{eq:QGci:sub}.

\paragraph{The incoherent to coherent quench} has the opposite signs of the couplings 
$g_{f}=-g_{i}=g>0$. Here we consider both combinations $F^f_{\Lambda,1}(t_f)F^{i}_{\Lambda,2}(t_i)$
and $F^f_{\Lambda,2}(t_f)F^{i}_{\Lambda,2}(t_i)$ since the first one is oscillating after the 
quench whereas the second one is purely decaying. Using 
$\Gamma_{1/2}^{*i}\approx 2^{-g} T_K^i$ and $\Gamma_{1/2}^{*f}\approx 2^{g}T_K^f$, we obtain
\begin{align}
\nonumber
\int_{0}^{\infty} F_{\Lambda,1}^{f}(t-t_q)F_{\Lambda,2}^{i}(t_q) d\Lambda &\approx 
-\left(2\cos(\Omega_{f}(t-t_{q}))-1\right)\\
&\hspace{-4cm}\times
\frac{e^{-\frac{1}{2}\Gamma_{2}^{*i} t_{q}-\Gamma_{1}^{*f} (t-t_{q})}}{t_{q}}\, 
\frac{2}{\sqrt{T_K^{*i}T_K^{*f}}}
\frac{\left(T_{K}^{i}t_{q}/2\right)^{-g/2}}{\left(T_{K}^{i}t_{q}/2\right)^{-g}-\frac{1}{2}},
\label{eq:F1F2:ic-c}
\end{align}
and
\begin{align}
\nonumber
\int_{0}^{\infty} F_{\Lambda,2}^{f}(t-t_q)F_{\Lambda,2}^{i}(t_q) d\Lambda &\approx
\frac{e^{-\frac{1}{2}\Gamma_{2}^{*i} t_{q}-\frac{1}{2}\Gamma_{2}^{*f} (t-t_{q})}}{t}\\
&\hspace{-4cm}\times
\frac{1}{\sqrt{T_K^{*i}T_K^{*f}}}
\frac{\left(T_{K}^{i}t/2\right)^{-g/2}}{\left(T_{K}^{i}t/2\right)^{-g}-\frac{1}{2}}
\frac{\left(T_{K}^{f}t/2\right)^{g/2}}{\left(T_{K}^{f}t/2\right)^{g}-\frac{1}{2}}.
\label{eq:F2F2:ic-c}
\end{align}
For $t-t_q\ll t_q$, we can replace $t$ by $t_q$ in the last equation. 
Furthermore, using Eq.~\eqref{eq:prelax_intermediate_t}, we can use for $T_K^i t_q\gg 1$
\begin{equation}
\label{eq:P(t_q)}
P^i(t_q)\,\approx\,g\,e^{-\frac{1}{2}\Gamma_2^{*i}t_q}\,\frac{(S^-_{T_K^i t_q})^2}{T_K^i t_q}\,.
\end{equation}
Therefore, the sum of the two contributions \eqref{eq:F1F2:ic-c} and \eqref{eq:F2F2:ic-c} 
multiplied by $-g_i=g$ (i.e. the contribution to $P(t)$ from the quench) can be written as
\begin{multline}
-g_i\,\int_{0}^{\infty} F_{\Lambda}^{f}(t-t_q)F_{\Lambda,2}^{i}(t_q) d\Lambda
\approx\\ \approx
 A \Biggl[
-\frac{2}{S^{-}_{T_{K}^{i}t_{q}}}
e^{-\Gamma_{1}^{*f} (t-t_{q})}( 2\cos(\Omega_{f}(t-t_{q})) -1 )
+\\+
\frac{S^{+}_{T_{K}^{f}t_{q}}}{S^{-}_{T_{K}^{i}t_{q}}}
e^{-\Gamma_{2}^{*f} (t-t_{q})/2}\Biggr]
P^i(t_{q}),
\end{multline}
where $A=\sqrt{T_{K}^{i}/T_{K}^{f}}=(\Gamma_{0}/\omega_c)^{g}<1$.
This leads straightforwardly to Eq.~\eqref{eq:QGic:sub}.


\begin{thebibliography}{99}
\bibitem{footnote_1}
The prefactor of $P_{\text{pole}}(t)$ agrees with the one of NIBA only to order $g^0$. 
We were informed by U.\ Weiss that computing it within improved NIBA \cite{Egger97} gives a result which agrees 
to ours to order $g$.
\end{thebibliography}
\end{document}